# Nanoelectromechanical Sensors based on Suspended 2D Materials

Max C. Lemme[1,2,*], Stefan Wagner,[2] Kangho Lee[3], Xuge Fan[4], Gerard Verbiest[5], Sebastian Wittmann[6], Sebastian Lukas[1], Robin J. Dolleman,[7] Frank Niklaus[4], Herre S. J. van der Zant[8], Georg S. Duesberg[3], Peter G. Steeneken[5,8]

[1]Chair of Electronic Devices, RWTH Aachen University, Otto-Blumenthal-Str. 2, 52074 Aachen, Germany

[2]AMO GmbH, Advanced Microelectronic Center Aachen (AMICA), Otto-Blumenthal-Str. 25, 52074 Aachen, Germany

[3]Insitute of Physics, Faculty of Electrical Engineering and Information Technology, Universität der Bundeswehr Munich, Werner-Heisenberg-Weg 39, 85577 Neubiberg, Germany

[4]Division of Micro and Nanosystems, KTH Royal Institute of Technology, Malvinas väg 10, 10044 Stockholm, Sweden

[5]Department of Precision and Microsystems Engineering, Delft University of Technology, Mekelweg 2, 2628 CD Delft, The Netherlands

[6]Infineon Technologies AG, Wernerwerkstraße 2, 93049 Regensburg, Germany

[7]2nd Institute of Physics, RWTH Aachen University, Otto-Blumenthal-Str., 52074 Aachen, Germany

[8]Kavli Institute of Nanoscience, Delft University of Technology, Lorentzweg 1, 2628 CJ Delft, The Netherlands

*Correspondence should be addressed to Max C. Lemme; max.lemme@eld.rwth-aachen.de

**Abstract** – The unique properties and atomic thickness of two-dimensional (2D) materials enables smaller and better nanoelectromechanical sensors with novel functionalities. During the last decade, many studies have successfully shown the feasibility of using suspended membranes of 2D materials in pressure sensors, microphones, accelerometers, mass and gas sensors. In this review we explain the different sensing concepts and give an overview of the relevant material properties, fabrication routes and device operation principles. Finally, we discuss sensor readout and integration methods and provide comparisons against the state-of-the-art to show both the challenges and promises of 2D material based nanoelectromechanical sensing.

## Introduction

Two-dimensional (2D) materials have excellent material properties for sensor applications due to their large surface-to-volume ratio and unique electrical, mechanical and optical properties **[1], [2]**.

More recently, the potential of 2D materials for sensing has been further extended by freely suspending 2D materials to form atomically thin membranes, ribbons or beams **[3]**–**[6]**. These types of suspended 2D material structures enable a new class of 2D suspended NEMS sensors, which is the focus of the present review. Suspending 2D materials eliminates substrate interactions, increases their thermal isolation and gives them freedom of motion, which opens a whole range of mechanical sensing modalities. In fact, many of the current micro- and nanoelectromechanical system (MEMS and NEMS) devices can be realized using suspended 2D materials, offering smaller dimensions, higher sensitivity and novel functionalities compared to their silicon-based MEMS and NEMS counterparts. This is because the performance and sensitivity of NEMS sensors often depends critically on the thickness of the suspended membrane or beam, which can reach its ultimate thinness when using suspended 2D materials. Moreover, new types of sensors can be enabled by exploiting the unique properties of 2D materials. Sensors in which the nanomechanical and/or electrical response of suspended 2D materials are used to sense environmental parameters, can be classified as 2D material NEMS sensors. Such 2D NEMS sensors therefore have the potential to provide novel and/or better solutions for applications such as the Internet of Things (IoT) or and autonomous mobility, which are expected to drive the demand for integrated and high-performance sensors for years to come.

Early studies investigated the application of graphene in NEMS as resonant structures **[7]**, which provide ultimate sensitivity for mass detection down to the hydrogen atom limit **[8]**. An overview of graphene-based nanoelectromechanical resonators was provided in a 2013 review paper **[9]** and the utilization of graphene and carbon nanotubes in NEMS was briefly summarized in Zang et al. **[10]**. However, it has recently become clear that graphene has potential for enabling a much wider range of NEMS sensors, with transition metal dichalcogenide (TMD) and 2D semiconductor materials also emerging in this application space **[6], [11], [12]**.

In this work, we present a review of 2D material NEMS sensors based on suspended graphene and related 2D materials operating in vacuum or gaseous environments. We discuss the relevant material properties, describe key fabrication technologies and evaluate the potential for Complementary Metal Oxide Semiconductor (CMOS) integration of 2D material NEMS sensors, specifically focusing on those topics relevant for these sensors that are not covered by previous reviews **[13]–[15]**. We present suitable transduction mechanisms that are of particular relevance to NEMS sensors and finally review the state-of-the-art in 2D membrane-based NEMS sensors applications, discussing pressure sensors, accelerometers, oscillators, resonant mass sensors, gas sensors, Hall-effect sensors and bolometers. This latter part of the paper is organized by application, not by material.

## Material properties of suspended 2D materials

In designing sensors and deciding on how to fabricate them, it is important to select a suitable 2D material. For that purpose, we discuss here the material properties that are relevant for nanoelectromechanical sensing. In fact, not all 2D materials are suitable to form suspended structures. As for graphene, many of its material properties are beneficial for forming freely suspended membranes, beams and ribbons, including chemical stability at atmospheric conditions, excellent mechanical robustness, stretchability of up to about 20 % **[16]**, a Young's modulus of 1 TPa **[17]**, intrinsic strength of 130 GPa **[17]**, room-temperature electron mobility of $2.5 \times 10^5$ cm$^2$/Vs **[18]**, excellent transparency, uniform optical absorption of $\approx$ 2.3 % in a wide wavelength range **[19]**, impermeability to gases **[20], [21]** (except hydrogen **[22]**) and the ability to sustain extremely high current densities **[23]**. Because graphene shows very strong adhesion to $SiO_2$ surfaces **[24]**, it can be suspended in one atom layer thick membranes that are mechanically stable **[25]**, and can be readily chemically functionalized **[26]**. However, it is important to point out that some of the extreme properties have been measured only in mechanically exfoliated, high-quality graphene samples that do not contain grain boundaries **[27]**, or for graphene on specific substrates such as hexagonal boron nitride **[18], [28]**.

Beyond graphene, other 2D materials also show promising properties for the use as membrane sensors, such as their relatively high in-plane stiffness and strength **[29]**. For instance, Young's moduli of monolayer h-BN, $MoS_2$, $WS_2$, $MoSe_2$, and multi-layer $WSe_2$ are reported to be 865 GPa, 270 GPa, 272 GPa, 177 GPa and 167 GPa, respectively **[29]**, in line with theoretical predictions **[30]**. Furthermore, the intrinsic strength of h-BN and $MoS_2$, two of the most studied 2D materials beyond graphene, are reported to be ~70.5 GPa and ~22

GPa, with fracture strains of 6-11 % and 17 %, respectively **[29]**, comparable to graphene. Hexagonal BN is an insulator that is used as a substrate and as encapsulation material for graphene and other 2D materials to improve their electronic transport properties **[28]** and mechanical stability. The piezoresistive gauge factors of monolayer $MoS_2$, bi-layer $MoS_2$ and $PtSe_2$ have been reported to be about -148 ± 19, -224 ± 19 and -84 ± 23 respectively **[6], [31]**, which are up to two orders of magnitude higher than commonly reported values in graphene with gauge factors (GF) between 2 and 6 **[25], [32]–[35]**. Therefore, compared to graphene, transition metal dichalcogenides (TMDs) offer piezoresistive readout of NEMS with much higher responsivity. Other 2D TMDs such as $WS_2$, $MoSe_2$ and $WSe_2$ are also predicted to have much higher piezoresistive gauge factors than graphene **[36], [37]**, emphasizing the potential of TMD-based piezoresistive membrane sensors. Table 1 compares the 2D material properties that are most relevant and interesting for applications based on suspended membranes, such as the Young's modulus, piezoresistive gauge factor and optical bandgap.

*Table 1: Comparison of the most relevant properties of suspended 2D materials. Reported results are obtained from experiments on suspended membranes, unless indicated otherwise between brackets.*

| | Young's Modulus (GPa) | Poisson's ratio | Fracture strain (%) | Mobility ($cm^2/Vs$) | Piezoresistive gauge factor | Optical bandgap (eV) |
|---|---|---|---|---|---|---|
| Highest-Quality Exfoliated Graphene | 800-1100 **[17], [38]** | 0.11-0.2 **[39]–[42]** | 0.3-30 **[17], [42]** | 200000 (suspended) **[43]** | 2-6 **[32]–[34]** | No bandgap |
| CVD polycrystalline Graphene | 1000 **[44]** | 0.13-0.2 **[39]–[41]** | 2 **[45]** | 350000 (supported) **[46]** | 2-6 **[32]–[34]** | No bandgap |
| h-BN | 223±16 **[47]** | 0.21 **[48]** | 17 **[49]** | dielectric | - | 5.9 **[50]** |
| $MoS_2$ | 270±100 **[51]** | 0.27 **[52]** | 6-11 **[53]** | 73 (supported) **[54]** | -148 ± 19 (monolayer) **[31]** -224 ± 19 (bilayer) **[31]** | 1.9 (monolayer); 1-1.6 (multilayer) **[55], [56]** |
| $MoSe_2$ | 177.2 **[57]** | 0.23 **[57]** | 2.55 **[57]** | - | 1800 (theory) **[58]** | 1.59 **[59]** |
| $PtSe_2$ | - | - | - | Mostly <15; 210 **[60]** | Up to -85±23 (few layer) **[6]** | 1.2-1.6 (monolayer); 0.2-0.8 (bilayer); none (multilayer)**[61], [62]** |
| $WS_2$ | 272 **[63]** | 0.21 **[64]** | - | 214 **[65]** | 14 **[37]** | 2 **[66]** |
| $WSe_2$ | 167.3 **[67]** | 0.19 **[64]** | 7.3 **[67]** | - | 3000 (theory) **[58]** | - |
| Black Phosphorus | 46-276 **[68]** | 0.4 **[68]** | 8-17 **[68]** | 10000 (supported) **[69]** | 69-460 **[70], [71]** | - |

The values in Table 1 are extracted from measurements at room temperature under application relevant conditions. Some properties like charge carrier mobility values have only partly been investigated for the suspended 2D materials. The terms “suspended” and “supported” therefore indicate how the value was obtained. In general, due to differences in fabrication and characterization procedures, large variations in the different material properties are found in literature, which leaves many open questions for NEMS device functionality. In addition, built-in stress in suspended 2D materials is generally large and difficult to control, while having a tangible influence on the static and dynamic characteristics of 2D material NEMS **[72]**. Built-in stress in fully-clamped graphene membranes can reach $10^2$ to $10^3$ MPa **[17], [20], [38], [73]–[78]** while stress in doubly-clamped graphene ribbons or beams can reach $10^1$ MPa **[7], [79]–[83]**, or about 200 MPa to 400 MPa in graphene ribbons with suspended silicon proof mass **[72]**. The built-in stress can substantially influence the resonance frequencies of resonators and accelerometers, as well as the force-induced deflection and strain in suspended 2D material membranes **[72]**. The fabrication process can further influence built-in stress, i.e., through design features, material growth and the transfer material **[73]**.

It should be noted that only a few of the materials listed in Table 1 have been shown to survive as self-suspended 2D material membrane, ribbon or beam structure **[3]–[5], [7]**, however many of these 2D materials may still be employed in NEMS sensors in form of multi-layers or in combination with more stable suspended support layers such as graphene to form suspended heterostructures **[63], [84], [85]**. 2D materials may also be combined with polymer layers to form suspended membranes and beams **[6], [86], [87]**. The buckling metrology method has been recently revisited as an alternative method to determine the Young’s modulus of 2D materials and generally results in comparable experimental values as conventional metrology methods (where available) **[88].**

## Fabrication methods for suspended 2D material devices

### *2D material exfoliation and growth*

Initially, manual exfoliation of flakes from bulk crystals was the most popular fabrication method in 2D material research because it results in single crystalline nanosheets with low defect density. Although the method enables the fundamental exploration of material properties and new device concepts, it is not a process that can be scaled up to high-volume production for mass market applications. An alternative method to obtain larger quantities of 2D material, is liquid-phase exfoliation in common solvents **[89]**. In this production method guest molecules or ionic species are intercalated between layers of bulk crystals, increasing the interlayer spacing and reducing binding, thus facilitating exfoliation of monolayers in subsequent processes, such as ultrasonication **[90]**, thermal shock **[91]** or shear **[92]**. Liquid exfoliation leads to dispersions of flakes that can be printed or sprayed onto substrates for sensor applications. This approach is suitable for example in applications, where the device functionality is mediated by mechanisms beyond the intrinsic material related to interfaces between the (randomly) oriented flakes arrangement, i.e. binding flake edges in gas and chemical sensors or current percolation between flakes in piezoresistive strain sensing **[93], [94]**.

In general, large-area chemical vapor deposited (CVD) graphene related materials are the preferred option for integrated NEMS sensors, because the method is in principle compatible with semiconductor technology **[13], [14]** and has the potential to result in uniform, reproducible layers. CVD graphene is typically deposited on a catalytic surface such as Cu or Ni, from which it can be transferred to arbitrary target substrates and the number of layers is precisely controllable **[95]–[99]**. Wirtz *et al.* managed to fabricate gas tight large area membranes (4 cm × 4 cm) by stacking 3 or more CVD grown graphene layers **[85]**. The properties of CVD graphene strongly depend on the material quality, the substrate material on which the graphene sheet is placed, and the crystal grain size, which typically is on the order of a few μm. Templated growth can lead to relatively large areas of crystalline CVD growth on copper **[100]** or sapphire wafers **[101]**, although full wafer scale of singly crystal growth has yet to be demonstrated. Despite the grain boundaries, CVD graphene is not always inferior to exfoliated "perfect" graphene, depending on the application case **[44], [102]**. Other available forms of graphene include epitaxial graphene grown on SiC substrates. CVD is also widely used to grow other 2D materials on a large-scale. A variety of different growth substrates are used depending on the targeted 2D material, for example Si/$SiO_2$, quartz, graphite or even other 2D material substrates for the growth of $MoS_2$, $WS_2$ or $WSe_2$ or metals such as copper, iron or platinum for the growth of h-BN **[85], [103]–[106]**. However, the field of large area synthesis of 2D materials is until evolving rapidly. For example, it is challenging to obtain continuous films and to control the thickness and quality is far from mature. An extensive overview of the production and process challenges has recently been presented in Backes *et al.* **[15]**.

An alternative synthesis approach introduced recently for transition-metal dichalcogenides (TMD) is thermally assisted conversion (TAC) utilizing vaporized chalcogenide precursors. For instance, Mo or more commonly $MoO_3$ can be converted to $MoS_2$ at high temperature **[107]–[112]**. This facile growth method is applicable to a wide range of TMDs, such as $MoSe_2$ **[113], [114]**, $WS_2$ **[115]–[117]**, $WSe_2$ **[118]**, $PtSe_2$ **[119]** or $PtTe_2$ **[120]**. The method yields continuous polycrystalline films, and therefore, pre-patterned transition-metals can be directly converted to structured TMDs. The thickness of converted TMDs is determined by the thickness of initial transition-metal layers. Thus, the TAC synthesis has advantages in terms of manufacturability of NEMS sensor devices.

*Fabrication of devices with suspended membranes*

There are several routes to fabricate devices with suspended membranes (often called "drums"), beams or ribbons of 2D materials. These routes can be distinguished by (1) the method of 2D material application (2D material transfer from the growth substrate to a target substrate (in contrast to 2D material growth on the target substrate) as shown in red color in Figure 1a-e) and (2) the method of creation of cavities below the membranes (etching underneath the 2D material in contrast to 2D material transfer onto a pre-etched cavity, as shown in green color in Figure 1f-j).

Figure 1a,b shows the option where the device substrates are fabricated before 2D material transfer. This includes the etching of cavities over which the 2D material is to be suspended, as well as the fabrication of electrical contacts, gate electrodes or sensing electrodes. Subsequently, 2D materials are transferred and suspended using wet transfer **[121]**, or dry transfer using PDMS stamps **[122]**, frame-based **[99], [122]–[125]** or other methods **[126]**, each with its advantages and disadvantages **[84]**. It should be noted that compared to conventional transfer, transfer of 2D materials over cavities is challenging. Stamp transfer (Figure 1f) can fail by delamination due to low adhesion forces, rupture of the membranes at cavity edges and stiction on the cavity bottom **[127]**. Alternatively, the transfer layer can be removed by etching (Figure 1g), which poses other challenges. The application of pressure on the stamp can affect the value and uniformity of the pre-tension in the suspended membrane and thus influence its mechanical resonance frequency and stiffness. Moreover, nonuniformity of the strain in the transfer layer can lead to wrinkled graphene membranes,

and polymeric residues of a few nanometer from the stamp can be present **[128]**. In general, few-layer membranes are more stable, show a higher yield of intact membranes after fabrication **[127]** and can be suspended across larger areas.

After the 2D material is successfully suspended using dry (Figure 1a,f) or wet (Figure 1b,g) transfer, it is important to minimize the impact of subsequent process steps to in order to reduce the risk of damaging the membrane and decreasing the yield of suspended 2D material membranes **[84]**. Process steps involving liquids suffer from capillary effects during drying and evaporation of the liquids, which typically decreases the yield of intact membranes **[84]**. Critical point drying (CPD) helps in this respect, but cannot be applied to membranes that seal holes because the high CPD pressures of more than 50 bar outside pressure, can break the membranes. Here, a "transfer last" method (Figure 1a,f) is an option to create sealed membranes as required for absolute or sealed gauge pressure sensors **[129]**. Another option is to seal the membrane at a later stage in the process **[21]**. Ribbons can be either structured on the growth substrate and then transferred with alignment routines **[130]** or have to be structured after suspension, which is technologically extremely challenging.

Some of the issues can be avoided by either growing **[131], [132]** or transferring unsuspended 2D materials directly on the device substrate **[72], [133]** (Figure 1c-e). It can then be patterned and subsequently the membrane can be released by isotropically under-etching (Figure 1h,i), by using a sacrificial layer **[134]–[137]** or by releasing the membranes from the backside (Figure 1j). The remaining through-hole can be left open or resealed after release **[133], [138]**. Process steps that avoid capillary forces during drying, such as CPD or hydrofluoric acid (HF) vapor etch can be used to avoid stiction and increase the yield of intact suspended membranes. Cleaning procedures for suspended 2D material devices are very delicate, because traditional methods used in MEMS manufacturing, such as ultra-sonic assisted dissolving or oxygen plasma ashing are aggressive towards suspended 2D materials and thus, these approaches are not suitable **[137]**.

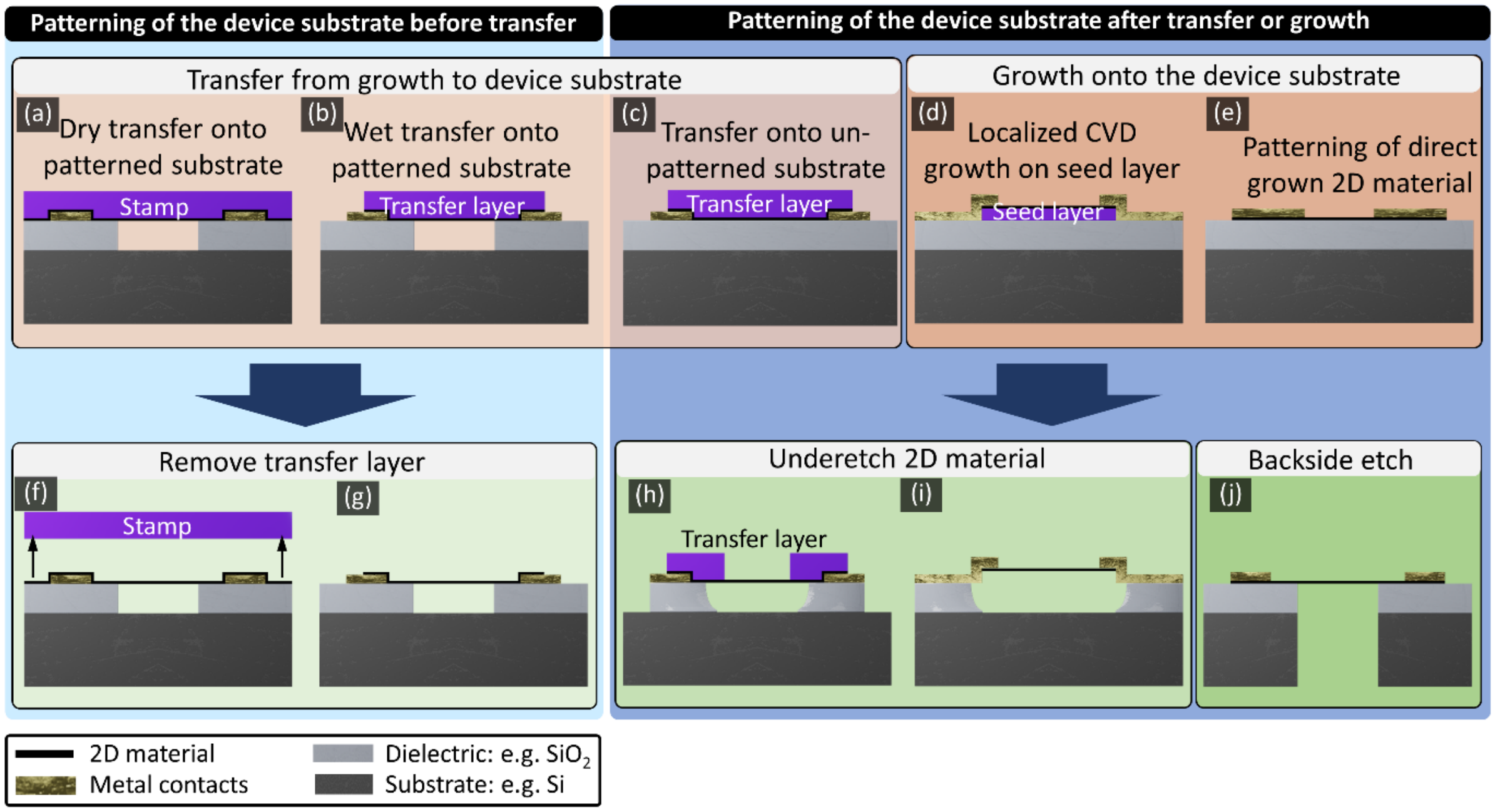

*Figure 1: 2D NEMS device fabrication methods: (a)-(e) Create a 2D material layer on the device substrate, where for (a) and (b) the device substrate is prepatterned and for (c)-(e) the substrate is patterned afterwards. (f)-(j) show post 2D material layer fabrication steps to create suspended membranes.*

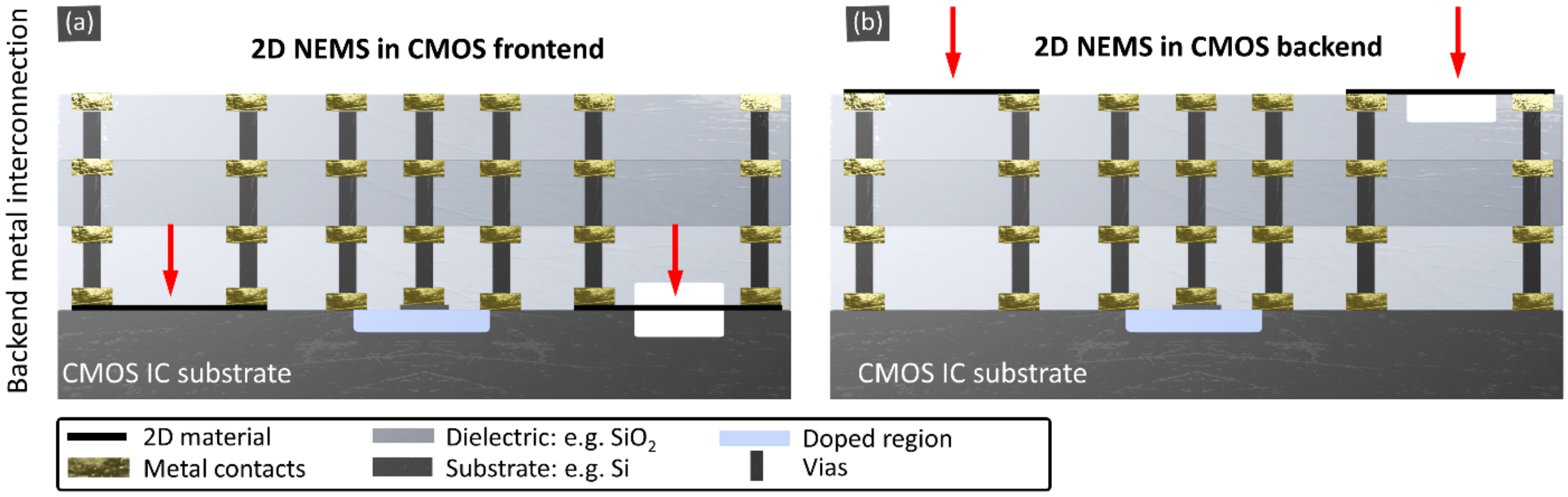

*Figure 2: CMOS integration of 2D NEMS sensors in backend (a): NEMS sensor devices integrated in the backend with interconnect layers stacked on top and frontend (b): integration of the 2D material in the frontend on top of the interconnect layers. The silicon IC substrate (dark grey), with transistors (blue) and interconnect metals (gray/yellow) are shown. Red arrows indicate the location of the black suspended graphene.*

## *CMOS integration*

Eventually it will become of interest to monolithically integrate suspended 2D materials with CMOS integrated circuits (ICs). Depending on the type of sensor and fabrication flow the sensor can be integrated both in the backend (Figure 2a) and in the frontend (Figure 2b) of the CMOS process. In both cases, devices with suspended 2D material membranes should be fabricated in a CMOS compatible way by growing the materials on wafer-sized substrates or by selective growth. The best process candidates are CVD and TAC, where the 2D material size is limited only by the reactor size. Wafer-scale transfer of graphene has been demonstrated and can in principle be integrated as a back-end-of-the-line process **[139]–[143]**. Direct growth of 2D materials in the back-end-of-the-line (Figure 2b) is only permitted if the growth temperature is below 450 °C, which is for example possible for $PtSe_2$ with a growth temperature of 400 °C or less **[119], [144]**. To realize CMOS integration, many challenges still need to be addressed. In particular front-end-of-the-line integration (Figure 2a) of suspended 2D materials is still very challenging **[13]**, because the material needs to survive all subsequent CMOS process steps. Besides realizing high-yield methods for the process steps discussed above, compatibility to CMOS temperature budgets, material interactions, delamination requirements, low contact resistances, packaging methods and reliability requirements will need to be dealt with.

Metrology is a general and ongoing challenge towards commercialization of 2D materials. This is augmented in membrane-based structures, scanning electron microscopy (SEM) is an option, but typically alters membrane properties due to the electron beam assisted deposition of hydrocarbon molecules. Raman spectroscopy is a non-invasive method if applied with low laser power, and can be extended to Raman Tomography **[145]**, which allows taking three-dimensional images of the entire device. Laser scanning microscopy is also feasible and non-invasive and can provide information about membrane deflection **[146]**. In addition, atomic force microscopy (AFM) **[147]** resonant interferometry **[148]** and colorimetry **[149]** can give useful information on the mechanical shape and stiffness of suspended 2D membranes.

## Readout and transduction mechanisms

A number of electrical transduction mechanisms can be utilized for readout of 2D material NEMS sensors. Although optical readout and analysis techniques **[7], [148]** are very convenient and useful for fundamental studies, we focus here on electrical readout techniques since they are more easily and seamlessly integrated for practical NEMS sensor devices.

The main electromechanical transduction and readout techniques suitable for 2D material NEMS sensors are piezoresistive readout, capacitive readout and transconductance readout. In addition, the electrical resistance of 2D material membranes can be used to sense changes in temperature, strain, carrier concentration or mobility that are induced by surface interactions (e.g. gas adhesion causes doping of the 2D material). It is important to note that the electrical resistance of 2D materials, especially graphene, is extremely sensitive to various environmental parameters, which means that parameters such as small changes in the air humidity **[150]–[153]**, light **[154], [155]**, gases **[119], [151], [152], [156]**, or temperature can strongly affect the electronic properties of a 2D material. Thus, for reliable use as sensors, these cross-sensitivity effects either have to be eliminated, by shielding or packaging, or they should be corrected for based on a calibration curve that eliminates environmental changes using input from a temperature or humidity sensor or reference device that is integrated in the same system **[6], [25]**. For resistance and Hall-voltage measurements of 2D material NEMS sensors, it is important to realize low contact resistances and use high-mobility graphene, a general topic that receives considerable attention **[157]–[162]**. In the following we now discuss the main electrical readout mechanisms of 2D sensors, piezoresistive, capacitive and transconductance readout.

Piezoresistive readout

The piezoresistive effect is defined as the change in electrical resistivity due to applied mechanical strain, which is related to the deflection of a membrane. The gauge factor (GF) is a measure for the piezoresistive effect **[163]**:

$$\boldsymbol{GF = \frac{\Delta R/R}{\Delta L/L} = \frac{\Delta R/R}{\varepsilon} = 1 + 2\nu + \frac{\Delta\rho/\rho}{\varepsilon}} \qquad (1)$$

It is defined as the ratio of the change in the electrical resistance ΔR to the change Δε= ΔL/L in mechanical strain (change in absolute length). The geometric deformation is described by the term 1+2ν, with ν as the Poisson's ratio. The gauge factor is directly related to the sensitivity of a piezoresistive sensor. Metals, such as constantan, which is used for commercial metal strain gauges, show a relatively low positive gauge factor of 2 **[164]**. Semiconductors, such as Si have a gauge factor of -100 to 200 **[165]**. 2D materials show piezoresistive properties as well. Graphene has a gauge factor between 2 and 6 **[25], [33], [34], [166]**, $PtSe_2$ up to -85 **[6], [144]** and $MoS_2$ between -148 and -40 for one to three layers **[31], [167]**. Simulations indicate a high gauge factor of up to 3000 for single layer $WSe_2$ **[58]** and around 1700 for single layer $MoSe_2$ **[58]**. These high values make piezoresistive readout an attractive method for readout of NEMS based on 2D materials. Moreover, piezoresistive readout can be scaled-down well **[168]**. Interestingly, for resonant strain gauges with nanoscale dimensions, such as doubly-clamped carbon nanotubes, silicon nanowires and graphene ribbons, the gauge factor of a strain gauge can be significantly amplified as a result of an asymmetric beam shape at rest **[72], [169]**.

*Capacitive readout*

Capacitive readout is an alternative method to determine the deflection of 2D membranes. For a deflection δ, the capacitance of a drum with area A and gap g is given by $C_{drum} = A\varepsilon_0/(g-\delta)$. The responsivity therefore scales as $dC/d\delta = A\varepsilon_0/g^2$ and increases by reducing the gap g. With respect to other deflection readout mechanisms, the important advantage of capacitive readout is that the capacitance only depends on the geometry of the structure, regardless of the membrane resistance and temperature. In practice however, it is difficult to fabricate membranes with gaps smaller than 100 nm with sufficient yield **[127]** without causing stiction during fabrication. Also, a small gap limits the maximum membrane deflection and thus, the maximum dynamic pressure range of the device. An alternative approach to increase responsivity is therefore to increase the area of the membranes, for instance by placing many graphene sensors in parallel **[87].** Another challenge is that there are usually parasitic parallel capacitances $C_{par}$ present between the top and bottom electrodes that need to be minimized to reduce power consumption and increase signal-to-noise ratio. This can be achieved by utilization of an insulating layer with a low dielectric constant and sufficient breakdown strength, a small overlap area between top and bottom electrodes (using local gates) and the utilization of an insulating, low dielectric constant substrate **[87]**. A unique feature of monolayer membranes, such as monolayer graphene with low carrier densities is that their capacitance is lowered by an effective series quantum capacitance **[170]**, especially close to the Dirac point. When a readout voltage Vg is applied across the sensor to determine its capacitance, this will not only affect the quantum capacitance, but can also result in an electrostatic pressure $P_{el} = \varepsilon_0 Vg^2/(g-\delta)^2$ that adds to the gas pressure and deflects the membrane. These effects need to be considered to accurately operate capacitive graphene pressure sensors, either by proper modeling, or by proper calibration.

*Transconductance readout*

Transconductance readout is a sensitive electrical readout method for 2D material membranes (see e.g. **[171], [172]**). It requires a three-terminal geometry, in which the conductivity of the 2D membrane is measured between a source and drain electrode, while a voltage is placed on a nearby gate electrode. When the membrane is deflected, the capacitance between gate and membrane changes and results in a different charge Q on the membrane ($Q = CV_g$), which results in a change in charge density and thus a different conductivity of the membrane, similar to that in the channel of a field-effect transistor.

*Readout of resonant sensors*

For resonant sensors, usually a vector network analyzer or spectrometer is used to determine the resonance frequency from a frequency spectrum or the transfer characteristic. In order to continuously monitor a resonance frequency, the resonant sensor can be configured in a direct feedback loop as a self-sustained oscillator that generates a signal with a sensor signal dependent frequency, that can for example simply be read-out by a digital frequency counter circuit that counts the number of zero-crossings per second. This method has been applied successfully to MEMS squeeze-film pressure sensors **[173]**. In more advanced implementations readout can be performed using phased locked loops **[174]**. Nevertheless, the feasibility of realizing an integrated portable resonant graphene sensor still needs to be proven.

*Actuation methods*

Actuation methods for 2D membranes include electrostatic actuation, opto- or electrothermal actuation **[21], [175]–[178]**, hydraulic pumping **[179]**, mechanical amplification **[180]** and piezoelectric excitation **[180], [181]**. In general, for realizing most types of sensors concepts, the challenge is more in the readout than in the actuation. Nevertheless, for sensors that utilize actuation voltages and currents, these need to be stable and noise-free, since any drift and noise at the actuation side will end up in the readout signal. The effects of noise can be mitigated by using a longer time-averaging, or by placing membranes in parallel to increase responsivity **[87], [182]**.

## Mechanical properties of suspended 2D material membranes and ribbons

2D material membranes and ribbons, specifically those made from graphene, can be made a factor 1000 thinner than those of current commercial MEMS sensor membranes or beams. As a consequence, these graphene membranes and ribbons have a much lower flexural rigidity. This allows either the reduction of the sensor size to only a few microns in diameter or side length while retaining the flexural softness of the membrane or beam, or a significant increase in sensor responsivity. However, to enable these, several challenges need to be tackled. The membrane/ribbon deflection needs to be determined with nanometer precision using accurate transduction mechanisms and the pretension $n_0$ in the graphene needs to be low enough to ensure that the responsivity is not limited by it. For the deflection of a doubly-clamped 2D material ribbon caused by a center point force, the deflection at the center of the ribbon is described by

$$\boldsymbol{F = 16\left[\frac{EWH^3}{L^3}\right]Z + 8\left[\frac{EWH}{L^3}\right]Z^3 + 4\left[\frac{T}{L}\right]Z} \qquad (2),$$

where F is the load applied at the center of the ribbon, Z the resulting deflection of the ribbon at its center (for large deflection with respect to the thickness of the ribbon), E the Young's modulus of the graphene, W the width of the ribbon, H the thickness of the ribbon, L the total length of the ribbon, and T the built-in tension force of the ribbon **[72]**. Another aspect of 2D material membranes and ribbons, that is intrinsically different from conventional devices is that the force-deflection curve of indentation experiments tends to become nonlinear at much smaller deflections than for bulk materials, due to the small thickness and high Young's modulus in graphene in combination with geometric nonlinearities (from the second term on the right-hand side of Equation 1) related to membrane stretching. This effect increases the stiffness, and reduces the sensor linearity, which in principle can be corrected by proper calibration. It will increase operation range, but reduces responsivity and will therefore require tradeoffs between dynamic range and responsivity **[182]**. Since graphene membranes and ribbons have a much smaller area, they feature higher thermomechanical 'Brownian motion' noise **[177]**, that translates for example for a circular membrane to a pressure noise $p_n$:

$$\boldsymbol{p_n^2 = \frac{4k_BT\omega_0 m_{eff}}{A^2Q}} \quad \left[\frac{\boldsymbol{Pa^2}}{\boldsymbol{Hz}}\right], \qquad (3)$$

where T is the temperature, Q the quality factor and $\omega_0$ the resonance frequency of the membrane. This equation shows that on the one hand 2D material pressure sensors have reduced noise due to their small effective mass $m_{eff}$, whereas on the other hand thermomechanical noise will increase as a consequence of their smaller area and higher resonance frequency. Nevertheless, it is often not the thermomechanical noise that limits NEMS sensor resolution in practice, but readout noise.

A further requirement on membrane properties in many NEMS sensors, such as in some pressure sensor, is that the membrane may need to be hermetically sealed, such that the pressure in the reference cavity is constant and gas leakage is negligible during its lifetime **[21]**. Despite the impermeability of graphene for gases **[20], [22]** it was found that gas can leak via the interface between the substrate and the graphene. This leakage path needs to be sealed for long-term pressure stability inside the reference cavity **[21]**. In pressure sensing applications, it is typically preferred to maintain a vacuum or a very low gas pressure environment in the cavity below the 2D material membrane, to avoid internal pressure variations with temperature according to the ideal gas law, or alternatively methods to correct for these using an integrated temperature sensor are required.

## 2D material NEMS sensors

### Pressure sensors

Silicon-based pressure sensors were the first microelectromechanical systems (MEMS) product to reach volume production **[184]**. The number of pressure sensors produced per year currently exceeds a billion units per year. Whereas the field of pressure sensing also includes liquid, tactile and touch sensing applications, we focus here on gas pressure sensors using suspended membranes, with main applications as altimeters, barometers, gas control and indoor navigation. MEMS pressure sensors usually determine the pressure from the pressure difference $\Delta p$ (see Equation 1) across a plate that induces a deflection $\delta = \alpha \Delta p A^2/t^3$, a geometry and material dependent factor $\alpha$.

Commercial MEMS sensors can resolve pressure differences as small as 1 Pa, corresponding to altitude changes of only 5 cm. To reach this resolution, an extremely low stiffness of the mechanical plate is required, resulting in diaphragm sizes of several hundreds of microns at membrane thicknesses in the order of 0.5-10 μm. In addition, highly sensitive membrane deflection detection circuitry is used, conventionally based on piezoresistive readout, but recently also capacitive readout, such as the SBC10 pressure sensor of Murata with a responsivity of 55 fF/kPa **[185]**. Reducing the size and improving the sensitivity of pressure sensors is generally of interest. For example, size may be a decisive form factor for wearable electronics. Enhanced sensitivity of 2D sensors may also enable new applications that are currently not feasible, like altimeters with sub-cm resolution for indoor navigation or pressure sensors for presence detection. Moreover, higher sensor sensitivity can reduce size, acquisition time, power consumption and cost of readout electronics.

In the following we will first discuss two types of static graphene pressure sensors: piezoresistive and capacitive pressure sensors. Then we will discuss two types of resonant pressure sensors and Pirani pressure sensors. Finally, we will compare the different types of pressure sensors.

*Piezoresistive pressure sensors*

The basic geometries and their operation principles of 2D piezoresistive pressure sensors are shown in Figure 3a-c and Figure 3d-f, respectively. The first subfigures (Figure 3a,h) show the device fabrication according to methods described in Figure 1 (coloring shows 2D material transfer or growth and method to suspend the membranes). When the membrane is bent by a pressure difference, it introduces strain into the material (Figure 3d-f) which is detected as a resistance change (Figure 3g). It is important to note that gasses or moisure that are in contact with the suspended 2D material membrane typically affect its resistance, which can interfere with the piezoresistive signal during pressure measurements **[25], [35], [151]**. In addition to self-suspended graphene membranes **[25]**, graphene resistors have been used to piezoresistively detect the motion of membranes made from SiN **[186]** or polymers **[187]**. Even though graphene enables very thin membranes, its piezoresistive gauge factor GF = $(\Delta R/R)/\varepsilon$ is relatively low (see Table 1**Fehler! Verweisquelle konnte nicht gefunden werden.**) **[35], [188]**. Other 2D materials have higher gauge factors (see Table 1) and are promising for improving piezoresistive pressure sensor sensitivity, as demonstrated for $PtSe_2$ **[6]**. The membrane area of graphene **[25]** and $PtSe_2$ **[6]** devices can be reduced to around 170 $\mu m^2$, which is significantly smaller than the area (90000 $\mu m^2$) of conventional MEMS pressure sensors **[172], [189]**. Low dimensional materials, such as carbon nanotubes **[190], [191]** or silicon nanowires **[10], [192]** can also be used for piezoresistive sensors, due to their high GFs **[193]**. However, these materials can only be used as sensing elements and usually need a separate membrane to support them, in contrast to 2D membranes that can have both a mechanical and electrical function. Such purely 2D material membranes combine a very thin membrane with the intrinsic readout mechanism and potentially enable up to four orders of magnitude smaller device footprints **[6], [25]**.

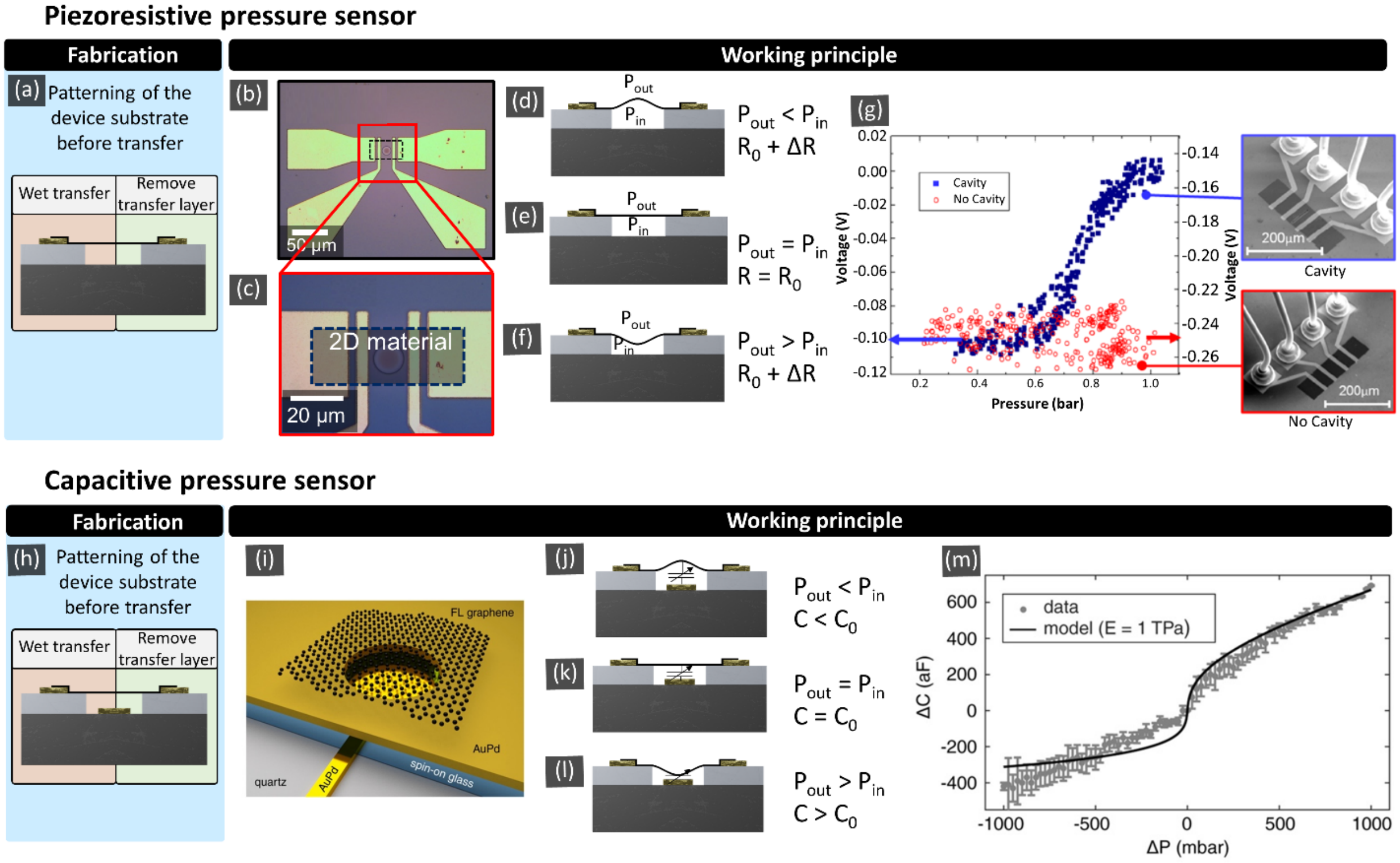

*Figure 3: Piezoresistive NEMS pressure sensor: (a) fabrication method of the suspended membrane (according to Figure 1), (b),(c) Example device image. [6]. (d)-(f) Working principle: pressure difference causes tension which alters the membrane resistance by the piezoresistive effect. (g) Graphene piezoresistive pressure sensor measurement [25]. Capacitive pressure sensor: (h) fabrication of the suspended membrane. (i) Device schematic [182], (j)-(l) Working principle: a pressure difference causes the membrane to deflect and alter the capacitance between the graphene and the bottom electrode. (m) Device measurement [182].*

### *Capacitive pressure sensors*

2D capacitive pressure sensors (Figure 3h,i) consist of a capacitor, which is formed between the membrane and a bottom electrode, such that a pressure change results in a capacitance change (Figure 3j-l). As can be seen in Fig. 3m, the capacitance is a nonlinear function of pressure. This is both due to the nonlinearity in the capacitance-deflection relation and due to the nonlinearity in the pressure-deflection curve (equation (3)). Main parameters that can influence the shape of this curve are the gap size, membrane thickness, Young's modulus, pre-tension, membrane radius and quantum capacitance. As can be seen from the slope of the curve in Figure 3m, the sensor is most sensitive when the pressure difference across it is zero.

When a capacitive pressure sensor is made out of a single graphene drum, its capacitance and change in capacitance is very small. For readout it requires detecting a small capacitance change on a large parasitic background capacitance. Even when using insulating quartz substrates to reduce the parasitic capacitance **[182]**, it is difficult to measure the capacitance changes, since responsivities of a drum with a 5 micron diameter are at most 0.1 aF/Pa, which at a voltage of 1.6 V corresponds to only 1 electron moving onto the graphene for a pressure change of 1 Pa. By utilizing a high frequency AC signal to charge and discharge the capacitor many cycles, signal to noise ratios can be improved to achieve a resolution of 2-4 aF/√Hz, requiring at least 20-40 of these drums in parallel to reach a pressure resolution of 1 Pa with an acquisition time of 1 second **[194]**. Recently, capacitive pressure sensors have been reported with many graphene drums in parallel that outperform the best commercial capacitive pressure sensors (SBC10 of Murata, responsivity 55 aF/Pa **[185]**), and that could be read out using a commercial IC **[195]**. With a large 5-layer graphene membrane a responsivity of 15 aF/Pa was reached **[196]** and an even higher responsivity of 123 aF/Pa was reached with graphene-polymer membranes **[87]**. Increasing drum diameter or further gap or tension reduction can also improve responsivity of graphene pressure sensors, although these options come with significant engineering challenges.

*Tension-induced resonant pressure sensors*

Resonant tension-induced pressure sensors monitor, similar to piezoresistive pressure sensors, the effect of gas pressure on the strain in a membrane. However, here the change in strain is monitored via its effect on the resonance frequency of the graphene membrane (Figure 4a,b). Bunch *et al.* **[20]** first utilized this effect to characterize the pressure difference across sealed graphene membranes in 2008. This demonstration of the extreme sensitivity of the resonance frequency to pressure was later confirmed with sealed graphene **[21]** and $MoS_2$ **[197]** membranes, resulting in variations in the fundamental resonance frequency of more than a factor of 4 (Figure 4c-f). A theoretical analysis of the dependence of the resonance frequency of a circular membrane on pressure found that the values of the Young's modulus that were extracted from the experimental fits are anomalously low **[21]**. It is still unclear whether this is related to wrinkling effects **[198]**, deviations from the theoretical shape and tension, or squeeze-film, slippage or delamination effects. Also, the pressure dependence of the quality factor of tensioned membranes is not fully understood **[136]** and might not only depend on the pressure difference, but also on the individual gas pressures below and above the membrane.

Typical responsivities d$\omega_0$/dp are larger than 200 Hz/Pa. It typically takes 1/200 second to determine a frequency change of 200 Hz, therefore this indicates that it might be possible to resolve pressure changes of 1 Pa in less than 5 ms. To actually achieve this, temperature **[176]**, mass loading and other effects that affect the resonance frequency of the membrane need to be prevented, or corrected with proper calibration using additional sensors. The low Q (Q of approximately 3) of graphene at atmospheric pressure will increase the power and time required to accurately determine the resonance frequency.

It should be emphasized that the high responsivity of tension induced pressure sensors can be attributed to the extreme thinness of graphene, which results in a low mass and thus in a very high initial resonance frequency $\omega_0$, but also in a relatively large strain and related tension-induced resonance frequency changes when the graphene “balloon” is inflated.

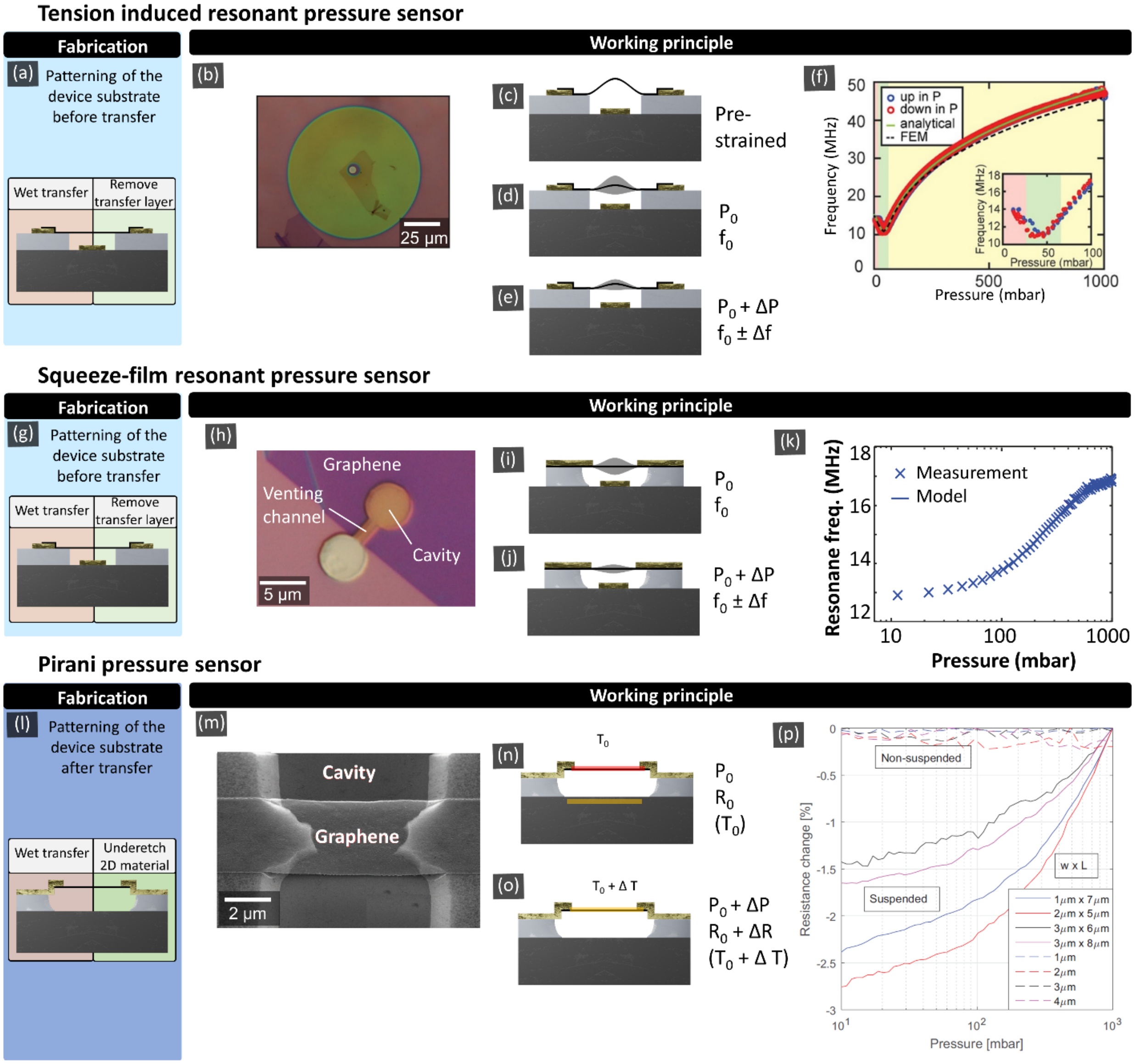

*Figure 4: Tension induced pressure sensor: (a) fabrication method of the suspended membrane (according to Figure 1), (b) example device [21]. (c)-(e) Working principle: the gas pressure difference across the membrane causes a membrane deflection and tension change that is measured via the resonance frequency. (f) Graphene tension induced pressure sensor measurement [21]. Squeeze-film pressure sensor: (g) fabrication of the suspended membrane, (h) example device [175]. (i), (j) Working principle: the stiffness and compressibility of the gas under the membrane increases the stiffness of the membrane that is measured via the mechanical resonance frequency. (k) Example measurement of a graphene-based squeeze-film pressure sensor [175]. Graphene Pirani pressure sensor: (l) fabrication of the suspended membrane; (m) example device of a Pirani pressure sensor [132]. (n), (o) Working principle: the temperature, and temperature dependent resistance, of the suspended, Joule heated graphene beam, depends on the pressure dependent gas cooling rate. (p) Example measurement of a Pirani pressure sensor based on graphene [132].*

*Squeeze-film resonant pressure sensors*

A second type of resonant pressure sensor is the squeeze-film pressure sensor. In contrast to the previously discussed sensors, squeeze-film pressure sensors do not require a hermetically sealed cavity (Figure 4g,h). The operation mechanism is based on the measurement of compressibility of gas inside the cavity under the graphene membrane. The compression occurs when the time it takes for pressure in the cavity to equilibrate is much longer than the period of the motion of the membrane, effectively trapping the gas in the cavity. It follows from the ideal gas law that the resonance frequency is $\omega_{res}^2 = \omega_0^2(\Delta P=0) + A P/(m g)$, where m is the membrane mass, so the low areal mass density of graphene is an advantage that increases the responsivity $\Delta \omega_{res}/\Delta P$ of the sensor. The change in the resonance frequency with respect to the vacuum value $\omega_0$ is dependent on the mass and geometry of the graphene cavity (Figure 4i,j). It has been shown **[175]** that the small graphene thickness and cavity depth result in a frequency change as large as 10-90 Hz/Pa, which is a factor of 5-45 higher than that in conventional MEMS squeeze film sensors despite the smaller area of the device (Figure 4k). More recently the feasibility of fabricating squeeze-film pressure sensors using transferless graphene (Figure 1d) has been demonstrated **[132]**.

*Pirani pressure sensors*

Pirani pressure sensors operate by measuring the pressure dependent thermal conductivity of the surrounding gas via its influence on the temperature dependent resistance of a suspended membrane (Figure 4l,m). In contrast to all other pressure sensors discussed above, the Pirani sensor does not mechanically move during operation. Conventionally, Pirani sensors are only used in vacuum systems: However, in **[199]** it was shown that the sensitivity range of these sensors can be brought to atmospheric pressure by reducing the gap down to 400 nm. The advantage of using graphene for Pirani sensors is that it takes much less power to heat a thin beam than a thick beam, and the temperature of the graphene beam depends more strongly on the cooling by surrounding gases due to its large surface to volume ratio (Figure 4n-p). With a transferless process flow (Figure 1d), the feasibility of graphene Pirani pressure sensors was recently demonstrated **[132]**. It should be noted that the response of Pirani pressure sensors is gas dependent, due to differences in thermal conductivity of different gases. This property might be employed to utilize the Pirani sensor as a gas sensor, when complemented by a pressure sensor that is independent of the type of gas.

*Pressure sensor comparison*

Important benchmark parameters for comparing different pressure sensors include size, power consumption, acquisition time, cross-sensitivity, reliability and production cost. In terms of performance, the capability to detect small pressure changes ΔP is an important parameter to compare the different sensors. To detect the signal of such a small change, it needs to be larger than the pressure noise in the system, i.e., the signal-to-noise-ratio SNR needs to exceed 1. Usually, the electrical readout noise (Johnson-Nyquist) is the dominant noise source that limits the SNR in these systems **[200]**. For a pressure change ΔP, the SNR is determined to compare the different types of pressure sensors (piezoresistive, capacitive and squeeze-film). The noise in a capacitive pressure sensor can be determined by using the charge noise of the capacitor $\boldsymbol{\sigma_Q = \sqrt{4\, k_B T C}}$ and the total energy costs for a measurement $E_{tot} = Pt_{readout} = NCV^2$, where $k_B$ is the Boltzmann constant, T the temperature, C the capacitance, P the electrical power consumption, $t_{readout}$ the readout time over which the measurement results are averaged, V the voltage and N the number of measurements **[200]**:

$$\boldsymbol{Noise = \sigma_C = \frac{\sqrt{4\, k_B T C/N}}{V} = C\sqrt{\frac{4k_B T}{Pt}}} \qquad (4)$$

The noise itself does not depend on the responsivity, but the capacitive signal dC = ΔP dC/dP does depends on the pressure change ΔP as well as the responsivity. By taking the ratio, the SNR can be calculated for the capacitive pressure sensor defined as:

$$\boldsymbol{SNR_{CAP} = \frac{1}{C_0}\frac{dC}{dP}\sqrt{\frac{P \cdot t_{readout}}{4k_B T}}\,\Delta P} \qquad (5)$$

Here, $C_0$ is the capacitance in the unloaded state. Note, that the minimum detectable pressure change corresponds to solving this equation for △P for SNR = 1. For comparison the SNR can be determined for a piezoresistive pressure sensor. An expression like (5) is found, with the term $1/C_0 \times dC/dP$ being replaced by $1/R_0 \times dR/dP$ for piezoresistive pressure sensors **[200]**. In case of the squeeze-film pressure sensor a factor Q needs to be added resulting in $1/C_0 \times dC/dP$ being replaced by $2/\omega_0 \times d\omega_{res}/dP \times Q$. We assume Q = 3 for graphene at atmospheric pressure **[201]**.

With these rough estimates of the SNR, based on an optimal performance of the readout system, different pressure sensors types can be directly compared to each other, which is shown in Figure 5. An SNR of $5.5 \times 10^{-6}$ $Pa^{-1}$ was calculated for both the $PtSe_2$ membrane-based piezoresistive by Wagner *et al.* **[6]** and the commercial capacitive pressure sensors Murata SCB10H **[185]**, which shows one of the highest SNR values available. The graphene membrane-based squeeze-film by Dolleman *et al.* **[175]** and capacitive pressure sensor by Davidovikj *et al.* **[182]** show values of $4.7 \times 10^{-6}$ $Pa^{-1}$ and $0.3 \times 10^{-6}$ $Pa^{-1}$, respectively. A SNR of $0.3 \times 10^{-6}$ $Pa^{-1}$ and $0.3 \times 10^{-7}$ $Pa^{-1}$ could be calculated for the piezoresistive graphene-based sensor by Wang *et al.* **[186]** and by Smith *et al.* **[25]**, respectively. These 2D material sensors were also compared to other low dimensional material-based NEMS pressure sensors (carbon nanotubes, Stampfer *et al.* **[190]**, silicon nanowires, Zhang *et al.* **[172]**) as well as to another commercial sensor, Epcos C35 **[202]** which is summarized in Figure 5. The $PtSe_2$ sensors show a factor of 5 to 200 higher SNR and up to 5 orders of magnitude smaller sensor area in comparison to state-of-the-art pressure sensors.

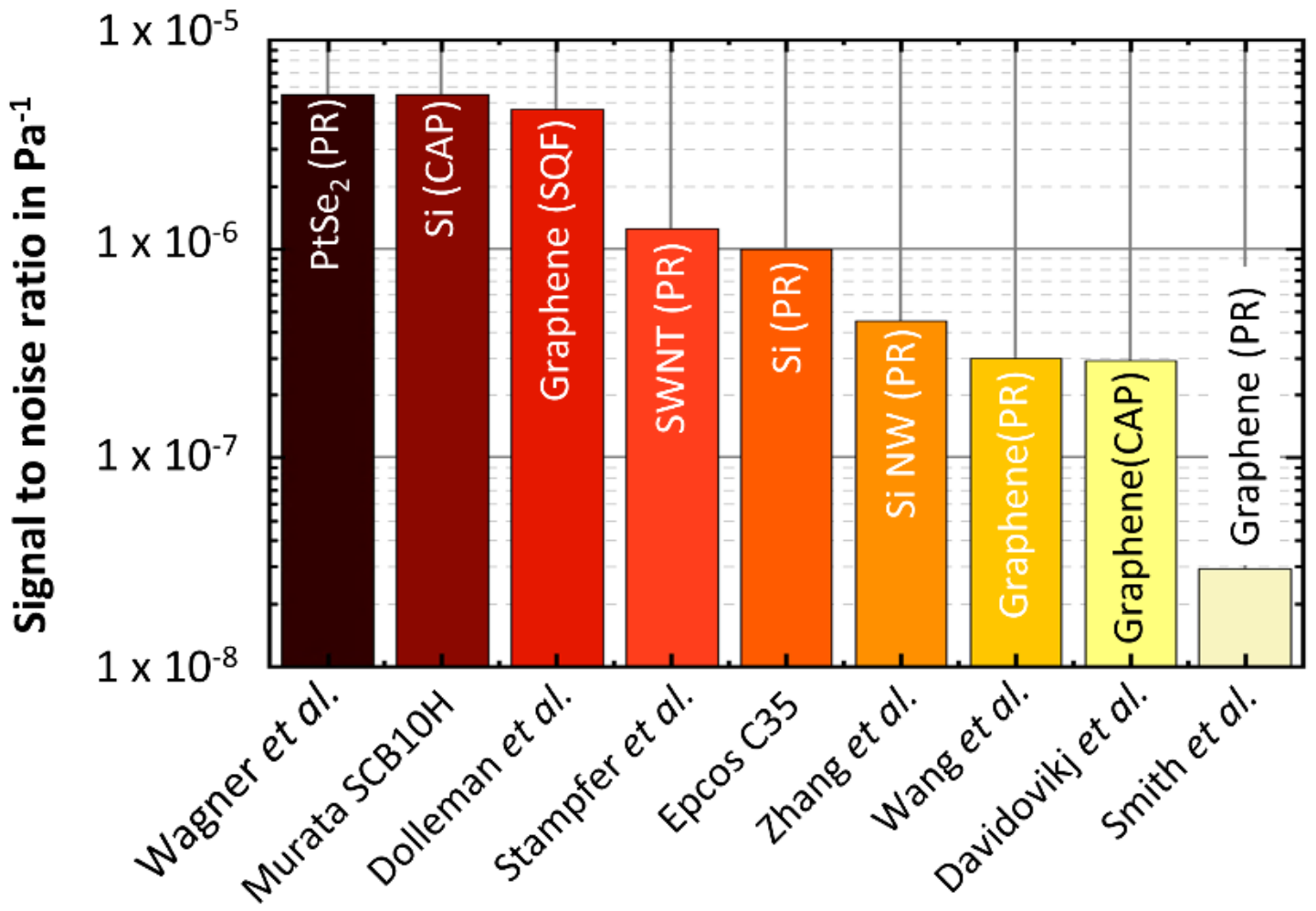

*Figure 5: SNR comparison of piezoresistive (PR), capacitive (CAP) and squeeze-film (SQF) MEMS pressure sensors. Included are: Wagner et al.* **[6]**, *Murata SCB10H* **[185]**, *Dolleman et al.* **[175]**, *Stampfer et al.* **[190]**, *Epcos C35* **[202]**, *Zhang et al.* **[172]**, *Wang et al.* **[186]**, *Davidovikj et al.* **[182]** *and Smith et al.* **[25]**.

### Graphene microphones

A microphone is essentially a pressure sensor that operates at audible or ultrasound frequencies. Similar to pressure sensors, the extreme thinness and the resulting flexibility of suspended 2D materials make them highly susceptible to sound pressure variations and thus suitable for application as microphones. In the last decades, MEMS microphones have replaced most conventional microphones in mobile devices and have become a billion-dollar market, where often multiple microphones are employed for realizing directionality and noise cancellation. The key advantage of using suspended graphene as a microphone membrane is its low stiffness $k_{eff}$. In conventional microphones, the stiffness cannot be lowered much further, because for a flatband frequency response it is required to have a resonance frequency $\omega_2 = k_{eff}/m_{eff}$ that exceeds the audible bandwidth (usually >20 kHz). Since graphene is extremely thin, it has a very small mass, allowing low stiffness to be combined with a high resonance frequency, offering interesting prospects for enabling wide bandwidth microphones that can detect small sound pressures. In addition, the low mass of graphene might be advantageous to reduce the pressure noise level based on equation (3). Besides improved performance, the advantages of graphene can also be utilized for area downscaling of microphones while maintaining current performance. This in turn can facilitate low-cost arrays of microphones that can enable directionality and might find applications in 3D ultrasound imaging and noise-cancellation. Challenges in reaching sufficient signal to noise ratio are even much tougher in microphones than in pressure sensors since current typical MEMS microphones boast responsivities (sensitivities) of >10 mV/Pa and impressive pressure noise levels below $p_n$ <10 μPa/√Hz **[203]**. This low-noise, high-responsivity performance has not yet been demonstrated with graphene membranes, but theoretically graphene is expected to outperform conventional MEMS membranes according to equation (3).

Condenser microphones with multilayer graphene membranes (20-100 nm thick) were reported with radii varying from 12 mm down to 40 μm **[146], [204], [205]**. These devices cover a frequency range from the audible domain **[204], [205]** up to the ultrasonic domain **[146]**. Devices with a small membrane diameter (Figure 6a-f) **[146]** operate over a wide frequency range that includes ultrasonic frequencies, while requiring low voltages, below the pull-in voltage of 1.78 V, which is well suited for use in mobile phones that provide a standard supply voltage of 2 V. Devices with a large membrane diameter **[204], [205]** require higher operation voltages but were also shown to function as a speaker. Importantly, some of the reported devices outperform high-end commercial nickel-based microphones over a significant part of the audio spectrum, with a larger than 10 dB enhancement of sensitivity, demonstrating the potential of graphene in microphone applications. Compared to conventional MEMS microphones with sensitivities of approximately -36 dB (around 15.8 mV/Pa), a supply voltage of 1.62-3.6 V **[206]** and an active membrane of 5 $mm^3$ **[207]** , graphene supported microphone diaphragms have sensitivities of up to 10 mV/Pa, at a supply voltage of 1 V and a diaphragm size of 38.22 $mm^3$ **[208]** . Thus, current silicon-based microphone technologies are even more sensitive than those using graphene, but microphone designs with two vibrating membranes are usually used to amplify the signal **[207]** , which is currently not the case with graphene.

**Ultrasound detection**

Recently, graphene-based high-frequency geophones have been introduced to detect ultrasonic waves in a silicon substrate **[181]** and to detect generalized Love waves in a polymer film (Figure 6g-j) **[209]**. In these works, a highly sensitive electronic read-out was employed reaching a resolution in ultrasonic vibration amplitude of 7 pm/√Hz. Interestingly, this resolution is independent of the mechanical resonance frequency of the suspended graphene membrane. The coupling mechanism between the substrate vibrations into the graphene membrane is currently still under debate, as the detected amplitudes are seemingly large. Recent work using an interferometric detection scheme suggests that graphene not just acts as a detector of the ultrasonic vibrations and resonant modes in the substrate, but also as an amplifier **[180]**. However, the physical origin of the strong coupling remains elusive. The possibility of using graphene for detecting vibrations or sound in solids could enable a new regime of ultrasound imaging at higher frequencies and smaller wavelengths than currently possible.

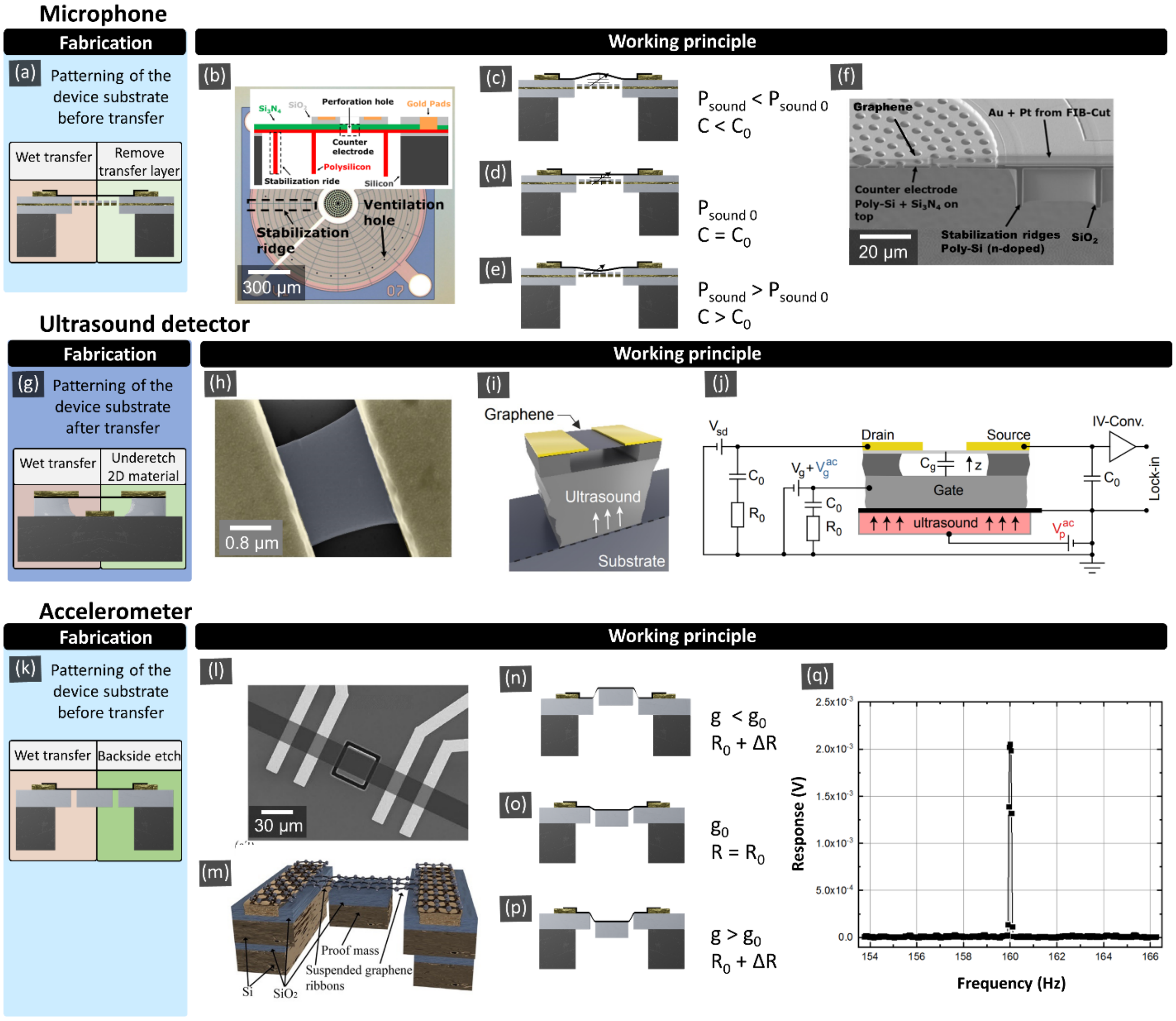

*Figure 6: Microphone: (a) fabrication method of the suspended membrane (according to Figure 1); (b),(f) images of an example device [146]. (c)-(e) Working principle: the sound pressure dependent deflection of the membrane is detected via its capacitance with respect to the backplate. Ultrasound detector: (a) fabrication of the suspended membrane; (h) example device [181] and (i),(j) working principle: the graphene membrane is moved by the ultrasound induced motion of its supports, and its motion is detected using transconductance readout. Accelerometer: (k) fabrication of the suspended membrane; (l),(m) example device [72], (n)-(p) working principle: the acceleration induced forces on the suspended mass cause tension in the graphene that is detected using the piezoresistive effect. (q) The output signal of an accelerometer [72].*

## Accelerometers

In current silicon-based MEMS accelerometers, the springs and interdigitated readout electrodes cause a significant increase in the device area. On the one hand this is caused by the requirement of a sufficiently small spring constant, which requires long compliant springs.

On the other hand, for capacitive readout MEMS accelerometers a sufficient capacitor area is required, which results in many interdigitated readout electrodes. Graphene and 2D materials on their own are not well suited for accelerometers, because their intrinsic mass is too small to achieve sufficient responsivity. 2D materials thus require an additional proof mass in the suspended region, which is displaced by acceleration forces. Although graphene has a small piezoresistive gauge factor, it can exhibit a large resistance change per Newton force ($1/F \times \Delta R/R$), because of its ultimate thinness. Its high Young's modulus and fracture strain further suggest that it is suitable for suspended devices with attached proof masses. Figure 6n-p shows an example of such a graphene NEMS accelerometer design, where the graphene simultaneously forms the springs of the spring-mass system and the piezoresistive transducer elements. The strain in the suspended graphene ribbons or membranes resulting from acceleration causes resistance changes in the graphene, due to the piezoresistive readout technique used in the accelerometers.

Double-layer graphene ribbons with large suspended silicon proof masses were realized with a conventional MEMS and NEMS manufacturing approach **[72]**. The graphene was suspended by dry etching followed by vapor HF etching to remove a sacrificial buried oxide layer (similar to Figure 1h). The suspended silicon proof masses had dimensions of up to $50\ \mu m \times 50\ \mu m \times 16.4\ \mu m$ (Figure 6k-m), which is more than three orders of magnitude heavier than the masses deposited on previous devices **[210]–[212]**. The graphene ribbons with suspended proof mass occupy at least two orders of magnitude smaller die areas than conventional state-of-the-art silicon accelerometers while keeping competitive sensitivity (Figure 6n-q) **[72]**. After normalization, the relative responsivity (resistance change per proof mass volume) in graphene ribbon accelerometers is at least one order of magnitude larger than the silicon state of the art. This demonstrates the potential to shrink the size of graphene-based NEMS accelerometers and gyroscopes despite graphene's low gauge factor.

The sensitivity of graphene accelerometers can be further improved by increasing the attached mass or by reducing the width of the suspended graphene **[72]**. From the perspective of material selection, the use of other two-dimensional materials like $MoS_2$ **[29], [31], [36]** or $PtSe_2$ **[6], [144]** with significantly higher piezoresistive gauge factors would also potentially improve the device sensitivity, although these materials need to be carefully evaluated with respect to their mechanical stability and adhesion force to the substrate. To this end, device designs based on fully clamped membranes improve the mechanical robustness by avoiding edges that are starting points for tearing under stress. However, this approach is a compromise as the signal response of fully clamped membranes is generally lower than that of ribbons with identical proof masses and trench width due to the lower strain levels  and parasitic parallel resistances **[133]**.

In addition to the above-mentioned demonstrations of graphene NEMS accelerometers, there are a limited number of experimental realizations of suspended graphene membranes or ribbons with attached proof masses. Micrometer-sized few-layer graphene cantilevers with diamond allotrope carbon weights fabricated by focused ion beam deposition have been used to study the mechanical properties of graphene **[210].** A kirigami pyramid was combined with cantilevers made of suspended graphene and supported 50 nm thick gold masses, but these devices had to be kept in liquid to maintain their mechanical integrity **[211]**. Finally, suspended graphene membranes circularly clamped by SU-8 that are supporting a mass made of either SU-8 or gold located at the center of the graphene membranes and that were evaluated as shock detector for ultra-high mechanical impacts **[212].** These reports utilized very small masses and some employed fabrication methods that are not considered compatible with semiconductor manufacturing. In addition, graphene-based resonant accelerometers have been proposed on theoretical grounds but not yet experimentally demonstrated **[213]–[215]**. In these concepts, the acceleration would act on suspended graphene beams or membranes, therby resulting in  added strain in the suspended graphene beams or membranes , thus causing a related shift in their resonance frequencies.

### Hall sensors

When a conductor, that is biased on one side, is exposed to an external magnetic field, charge carriers experience a Lorentz force that drives them in a direction perpendicular to the electric field and the external magnetic field. The resulting Hall voltage is a measure of the magnetic field and is proportional to 1/n, where n is the charge carrier concentration. The electronic structure of single layer graphene results in a very low carrier density at the minimum of its conductivity and thus high Hall voltage. In addition, the charge carrier concentration can be tuned to reach high responsivity. The ultimate signal to noise ratio of Hall sensors is proportional to the mobility μ. The very low effective mass of charge carriers in graphene translates into very high mobility at room temperature, which enables high-performance graphene-based magnetic field sensors. The mobility in graphene depends to a large extend on the (dielectric) environment, i.e. the interface with its surroundings. Relevant to this review, high mobilities of up to μ = 200,000 $cm^2$/Vs have been measured in suspended graphene **[216]–[218]**, which are significantly higher compared to up to μ = 20,000 $cm^2$/Vs for supported graphene on a $SiO_2$ substrate **[219]**. Suspended graphene Hall sensors are of interest (Figure 7a-d) because the voltage sensitivity SV of linear Hall sensors depends on the charge carrier mobility μ (SV ∝ μ·(W/L)), where W and L are the width and length of the device **[220]**. The carrier mobility of electrons is about 1,241 $cm^2$/Vs in silicon at a dopant concentration of approximately 1017 $cm^{-3}$ at room temperature **[221]**. The intrinsic SV is thus approximately 160 times greater for suspended graphene (at μ = 200,000 $cm^2$/Vs) than for silicon. Also, graphene shows a linear Hall response over several hundred mT **[222]** and surpasses commercial Hall sensors based on silicon technology **[223]**. Nevertheless, commercial monolithic silicon Hall sensors produced with BiCMOS technology, such as the Infineon linear Hall sensor series TLE499x **[224]** reach sensitivities up to 300 mV/mT at an operation voltage of 5.5 V and an operation range of ±200 mT. These high values are achieved through of the use of integrated amplifier circuits and enhance the intrinsic Hall effect in silicon. Such established integration technology is still missing for graphene, but improvements may be expected as the technology matures **[13], [14]**. Recent results indicate that graphene mobilities can be quite high when encapsulating graphene by Al2O3 **[220]**, hBN **[18], [46]** and $WSe_2$ **[225]**. This may be a promising route to also improve the performance of Hall sensors based on non-suspended graphene **[226], [227]**, which may be preferred for most applications, as it removes some of the fabrication challenges of suspended graphene membranes **[146]**. As discussed, the Hall effect provides an accurate method to detect the carrier concentration n. Suspended graphene Hall sensors, where the membrane is exposed to the environment, are thus promising as gas sensors, where molecules adsorbed to the graphene change its doping (= carrier density). Such sensors could be sensitive down to the single molecule level **[1]**.

## Gas Sensors

*Resistive gas sensors*

2D material gas sensors can be used for environmental monitoring **[12]**. These are generally based on the adsorption of analytes such as $NH_3$, $CO_2$, $H_2O$, and $NO_2$ on the sensor surface **[1], [150], [228]–[230]**. This is in contrast to conventional metal oxide gas sensors based on zinc oxide (ZnO) or tin oxide ($SnO_2$), that utilize surface reactions between oxygen and analyte molecules at grain boundaries. In 2D material gas sensors, the absorbed gas molecules induce charge carriers, that cause an electrical resistance change in the sensor (chemiresistor) (Figure 7g,h,k). Graphene chemiresistors are among the most investigated structures due their simple fabrication, characterization and miniaturization **[150], [231]–[236]**, as well as potential use for bio-sensors **[237]**. In a so-called chemical field effect transistor (ChemFET) **[1], [238], [239]**, the channel carrier concentration and conductance are modulated by applying a gate voltage to optimize gas sensing performance. Single layer graphene and 2D materials have the substantial advantage of an inherent large surface area-to-volume ratio, but can also exhibit low Johnson-Nyquist noise **[1]** and $1/f_{noise}$ **[240], [241]**. This unique combination can result in very high signal-to-noise ratios and potentially lower detection limits towards the individual gas molecule level. Suspending the channel effectively doubles the available surface area, and thus the achievable responsivity. In contrast, commercial chemiresistive gas sensors use e.g. metal-oxide sensor materials, because they are very sensitive to multiple gases, but require high operation temperatures of 150 ° C **[242]**, which are not needed in 2D material based chemiresistive gas sensors. Also, the measurable concentration range of commercial gas sensors is limited, because they saturate at high gas concentrations **[242]**. This limitation is less evident in 2D materials **[243]**. 2D materials haven been demonstrated with relative changes in resistance at room temperature of 39 % at 200 ppm $NO_2$ in air for graphene **[244]**, 10 % at 100 ppm $NO_2$ in $N_2$ for $MoS_2$ **[245]** and 0.25 % at 1 ppm $NO_2$ in $N_2$ for $PtSe_2$ **[119]**. Suspended bilayer graphene was used to measure $CO_2$ with high sensitivity (Figure 7f) **[246]**. MEMS MOS gas sensors based on silicon CMOS technology show resistivity changes from a few percent up to almost 100 % for different target gases, but at operating temperatures of 300 ° C **[247]**. This results in high-power consumption of the sensors and thus limits their suitability for low power applications such as smartphones.

Unfunctionalized suspended graphene resistors can also be used as gas sensors by measuring the thermal conductivity of a gas. A promising approach for improving response time and recovery time of indoor air quality sensors was demonstrated in **[248]**, where resistive graphene-oxide humidity sensors have been suspended on MEMS micro hotplates and characterized using a temperature modulation procedure. Schottky barrier diodes have been demonstrated to be extremely sensitive gas sensors, in which the Schottky barrier height (SBH) depends on analyte exposure, which in turn modulates electrical currents. Kim *et al.*

**[249]** proposed the effect of doping by liquid aromatic molecules on the SBH and Schottky diode ideality factor and Singh *et al.* demonstrated SBH modulation leading to a wide tunability of gaseous molecular detection sensitivity **[250]**.

Although graphene gas sensors can be very sensitive, a challenge is to make them selective, since they often respond to many different gases and other parameters, which is similar to metal oxide sensors. Selectivity can be achieved through dedicated functionalization layers that enhance the reactivity only for certain gases. In addition to graphene, 2D materials such as $MoS_2$ **[112], [251]**, molybdenum diselenide ($MoSe_2$) **[110], [252]**, molybdenum ditelluride ($MoTe_2$) **[253]**, tungsten diselenide ($WS_2$) **[117]**, niobium diselenide ($NbS_2$)**[254]**, rhenium disulfide ($ReS_2$) **[255]** or platinum diselenide ($PtSe_2$) **[119]**, have been shown to possess high gas and chemical sensor performance. Some TMD materials even show quite specific sensing behavior, in particular $PtSe_2$ has been shown to have a high selectivity towards $NO_2$, which also was validated theoretically **[119].** This may be exploited to enhance the sensitivity and selectivity through combining individual TMD sensors into sensor arrays **[256].** Such sensor arrays, functionalized or unfunctionalized, can then be combined into an electronic nose **[257].** Again, suspending these sensors will enhance the surface area and sensitivity, albeit at the cost of more challenging fabrication schemes, so that one has to choose an optimum cost/performance scenario.

Finally, repeatability and drift of gas sensors is a major general challenge, since the chemical binding energy of the gas molecules to the 2D material needs to be paid to remove the molecules and restore the sensor to its initial state. If the binding energy is close to $k_BT$ this might be performed by heating, otherwise light can be used to decrease recovery times.

*Permeation based gas sensing*

During the last decade several works have demonstrated the feasibility of fast molecular sieving in gases and liquids using membranes made of 2D materials **[258]–[260]**. It was shown that pores with sub-1nm diameters in these membranes can selectively sieve molecules or ions based on their molecular kinetic diameter. Specifically, it was shown **[258]** that small molecules such as $H_2$ and $CO_2$ permeate the membranes by a factor 1,000 faster than argon, nitrogen and methane gas. This methodology can also be used for permeation based gas sensing, as was shown in **[261]** where a change in gas composition caused an osmotic pressure across a graphene membrane. This pressure is a consequence of the permeability differences of the different gases, that effectively resulted in the graphene acting as a semi-permeable membrane. For even larger pore sizes, when going from molecular-sieving to effusion dominated permeation, these sensing principles can be utilized for gas sensing **[262]**, although with lower selectivity.

### Graphene mass sensors

The low mass of graphene makes it an interesting candidate for accurate mass sensing. Such a sensor, shown in Figure 7l-o, determines a mass change of the membrane or ribbon by monitoring changes in its resonance frequency. The mass change can be introduced by adsorbed or attached atoms or molecules on the surface of the membrane. The responsivity of resonant mass sensors is given by $\Delta\omega_{res} = -\frac{1}{2}\omega_{res}\,\Delta m/m_{eff}$ **[263], [264]**, which shows that for a small mass m of the graphene membrane or ribbon, a relatively large frequency shift will occur. The high sensitivity of this principle was shown by adding and removing layers of pentacene with an equivalent mass of 6 layers of monolayer graphene and monitoring its effect on the resonance frequency of a graphene membrane (Figure 7p) **[128]**. Such suspended graphene resonant mass sensors are expected to find applications in fields where it is required to determine mass changes much less than a monolayer of a 2D material. In comparison, conventional quartz crystal monitors have been shown to be able to measure the mass of a single monolayer of graphene **[128]**. The sensitivity of graphene based mass sensors can reach a value of $10^{-27}$ g/Hz **[265]** , which greatly outperforms silicon membrane based sensors, with typical sensitivity values of only $10^{-18}$ g/Hz **[266]**. Commercial mass sensors have even lower sensitivity values of around $60\times10^{-9}$ g/Hz **[267]**. In the ultimate limit, graphene nano-membranes with diameters of below 10 nm, which often occur naturally in graphene on silicon oxide substrate, have been theoretically predicted to be able to detect one hydrogen atom of mass, which would lead to a relative resonance frequency shift of $10^{-4}$.

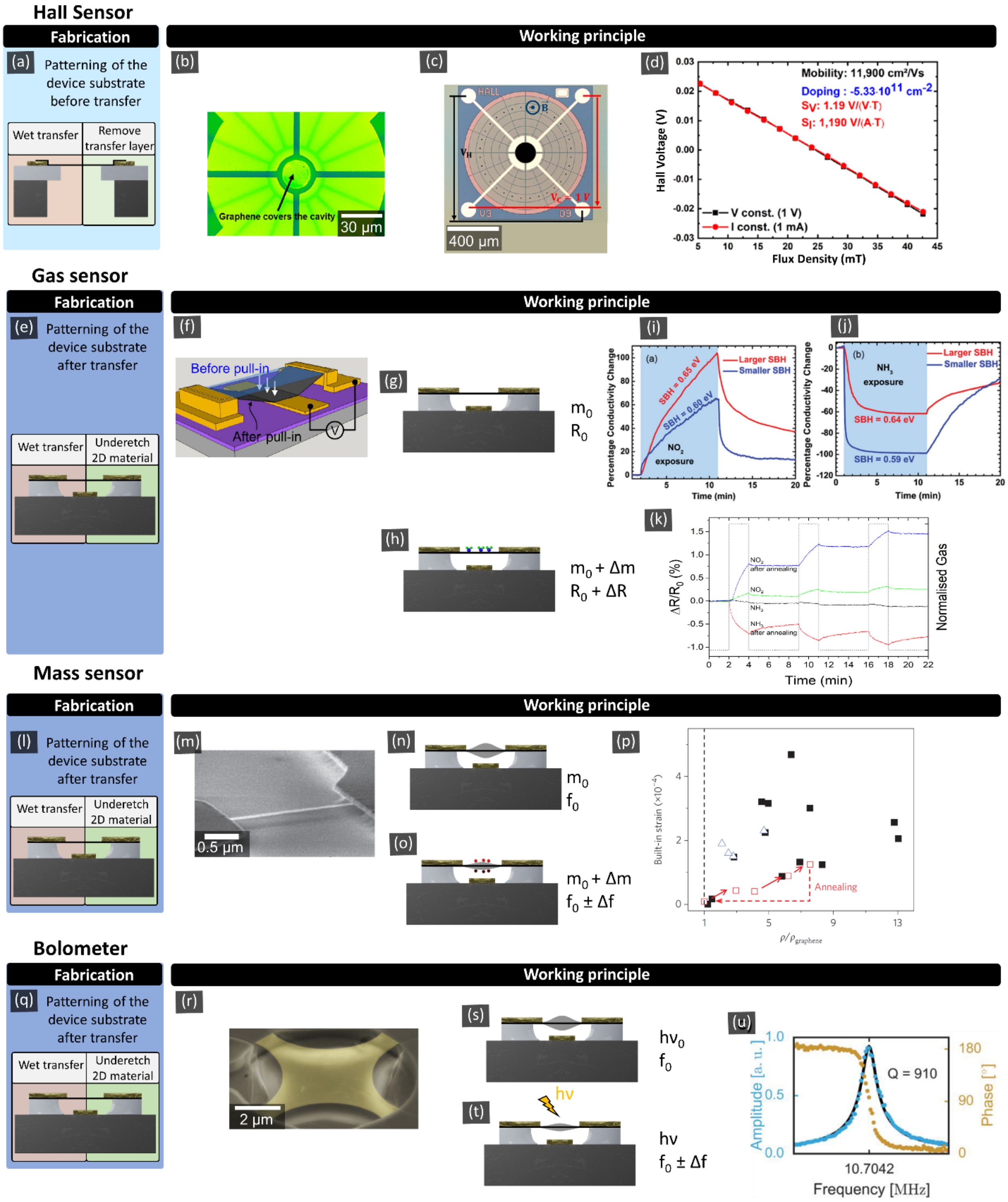

*Figure 7: Hall Sensor: (a) fabrication method of the suspended membrane (according to Figure 1), (b),(c) example device [146] and readout of an example device [146]. Gas sensor: (e) fabrication of the suspended membrane and ribbon, (f)example device [246]. (g), (h) Working principle: gas molecules adhere to the (functionalized) 2D material and alter its resistance via electronic or chemical interactions. (i),(j) Readout of an example device [250], (k) typical sensor response plot of $MoSe_2$ sensors depending on electron-donating/withdrawing gas [110]. Mass sensor: (l) fabrication of the suspended membrane, (m) example device, (n), (o) working principle: by measuring the resonance frequency the mass change of the membrane is derived. (p) Extracted mass and tension of the membrane during multiple loading cycles:[83]. Bolometer: (q) fabrication of the suspended membrane and*

*ribbon, (r) example device [268]. (s), (t) Working principle: when radiation heats the membrane, this alters its tension and causes a shift in mechanical resonance frequency. (u) Readout of an example device with a graphene membrane [268].*

## Graphene Bolometers

Bolometers are devices to detect absorption of electromagnetic radiation and light by monitoring the resulting temperature changes in a material via changes in its electrical resistivity. Especially for long wavelength infrared and THz radiation, bolometers are of interest, since there are few alternative detectors available in this frequency regime. At room temperature, where superconducting bolometers cannot be realized, suspended graphene is an interesting material for utilization of low-cost bolometers due to its ultra-wideband electromagnetic absorption and low heat capacitance due to its atomic thickness (Figure 7q-u). The high thermal conductivity and low temperature coefficient of resistance of graphene are drawbacks that have recently been mitigated by instead utilizing a resonant readout mechanism in a focused ion-beam structured suspended graphene bolometer (Figure 7q) **[268]**. However, cross-sensitivity to other signals (e.g. thermoelectric and photoelectric) needs to be also dealt with. Graphene-based resonant radiation detectors for the infrared range show a noise equivalent power of about 2 pW/Hz at room temperature **[268]**, and are thus in the upper range of conventional infrared bolometers based on vanadium oxide or nickel (1-10 pW/Hz) **[269]–[273]** .

There are many other types of 2D material based photosensors, but they are usually not suspended and fall therefore outside the scope of this review.

## Discussion and conclusions

While the field of silicon-based MEMS sensors is getting mature, the advent and discovery of 2D materials has brought us a set of nanomaterials for realizing novel NEMS sensors. Not only are these new materials thinner than any currently available CMOS or MEMS material, allowing drastic reductions of device size and enhanced sensitivity, there is also a larger range of materials emerging with exceptional properties. This large range of available material properties increases the freedom to engineer desired sensor properties for a particular application and to maximize sensitivity and reduce dimensions of the NEMS sensors. Moreover, by creating heterostructures of 2D materials, an even larger number of parameters will become available to optimize the sensor's electrical, mechanical, thermal, optical, chemical and magnetic properties. The possibilities are expanding even further, since new types of ultrathin materials for NEMS applications continue to emerge, like those based on complex oxides **[274]** and 2D organic magnetic membranes **[275]**.

In this review we have given an overview of the NEMS sensors and proof-of-concept devices based on suspended 2D materials that have been demonstrated during the last decade. These devices are almost always smaller than their conventional MEMS counterparts. Moreover, they show improved performance and sometimes even completely novel functionalities. Despite these successes there are still enormous challenges ahead to demonstrate that 2D material-based NEMS sensors can outperform conventional devices on all important aspects. One of these tasks is the establishment of high-yield manufacturing capabilities **[15]**. We have given an overview and comparison of the different potential fabrication routes and their challenges, focusing on the challenges related to suspended sensors. In this respect, the recent EU experimental pilot line is expected to set a big step towards high quality, high-volume graphene devices **[276]**. Of course, a platform approach where multiple types of suspended sensors can be produced in a single production flow is desirable, but it remains to be seen to what extent this can be realized. Other remaining tasks are sensitive and customized electronic sensor readout circuits, packaging and reliability testing for the 2D material NEMS sensors.

We believe that of all potential electronics applications for 2D materials, sensors made from non-suspended 2D materials could be one of the first to become commercially available. Suspending the materials inherently adds process complexity and challenges, and hence will likely take a longer time. Nevertheless, we are optimistic that, with joint efforts from both academia and industry, the first NEMS sensors based on 2D materials could hit the markets before the start of the next decade. In addition, 2D materials are now discussed for ultimate CMOS logic as stacked nanosheet transistors . This may trigger enormous, game-changing investments by industry, that would upend any predictions made by us today.

**Acknowledgement**

**Author Contributions:**

All authors contributed equally to the writing of the manuscript.

**Funding:**

This work was financially supported by the European Commission under the project Graphene Flagship [785219, 881603] and ULISSES [825272], the German Ministry of Education and Research (BMBF) under the project GIMMIK [03XP0210] and NobleNEMS [16ES1121], the German Federal Ministry for Economic Affairs and Energy (BMWi) and the European Social Fund in Germany under the project AachenCarbon [03EFLNW199], the Swedish Research Foundation (VR) (2015-05112), the FLAG-ERA project CO2DETECT funded by VINNOVA (2017-05108),the Dutch 4 TU Federation project High Tech for a Sustainable Future and the FLAG-ERA project 2DNEMS funded by the Swedish Research Foundation (VR) (2019-03412) and the German Research Foundation (DFG) [LE 2441/11-1].

## References:

[1] F. Schedin *et al.*, "Detection of individual gas molecules adsorbed on graphene," *Nat. Mater.*, vol. 6, no. 9, pp. 652–655, 2007, doi: 10.1038/nmat1967.

[2] E. W. Hill, A. Vijayaragahvan, and K. Novoselov, "Graphene Sensors," *IEEE Sens. J.*, vol. 11, no. 12, pp. 3161–3170, Dec. 2011, doi: 10.1109/JSEN.2011.2167608.

[3] N. Morell *et al.*, "High Quality Factor Mechanical Resonators Based on WSe2 Monolayers," *Nano Lett.*, vol. 16, no. 8, pp. 5102–5108, Aug. 2016, doi: 10.1021/acs.nanolett.6b02038.

[4] J. Lee, Z. Wang, K. He, R. Yang, J. Shan, and P. X.-L. Feng, "Electrically tunable single- and few-layer MoS2 nanoelectromechanical systems with broad dynamic range," *Sci. Adv.*, vol. 4, no. 3, p. eaao6653, Mar. 2018, doi: 10.1126/sciadv.aao6653.

[5] M. Will *et al.*, "High Quality Factor Graphene-Based Two-Dimensional Heterostructure Mechanical Resonator," *Nano Lett.*, vol. 17, no. 10, pp. 5950–5955, Oct. 2017, doi: 10.1021/acs.nanolett.7b01845.

[6] S. Wagner *et al.*, "Highly Sensitive Electromechanical Piezoresistive Pressure Sensors Based on Large-Area Layered PtSe2 Films," *Nano Lett.*, vol. 18, no. 6, pp. 3738–3745, Jun. 2018, doi: 10.1021/acs.nanolett.8b00928.

[7] J. S. Bunch *et al.*, "Electromechanical Resonators from Graphene Sheets," *Science*, vol. 315, no. 5811, pp. 490–493, Jan. 2007, doi: 10.1126/science.1136836.

[8] T. Mashoff *et al.*, "Bistability and Oscillatory Motion of Natural Nanomembranes Appearing within Monolayer Graphene on Silicon Dioxide," *Nano Lett.*, vol. 10, no. 2, pp. 461–465, Feb. 2010, doi: 10.1021/nl903133w.

[9] C. Chen and J. Hone, "Graphene nanoelectromechanical systems," *Proc. IEEE*, vol. 101, no. 7, pp. 1766–1779, Jul. 2013, doi: 10.1109/JPROC.2013.2253291.

[10] X. Zang, Q. Zhou, J. Chang, Y. Liu, and L. Lin, "Graphene and carbon nanotube (CNT) in MEMS/NEMS applications," *Microelectron. Eng.*, vol. 132, pp. 192–206, Jan. 2015, doi: 10.1016/j.mee.2014.10.023.

[11] B. Chitara and A. Ya'akobovitz, "High-frequency electromechanical resonators based on thin GaTe," *Nanotechnology*, vol. 28, no. 42, p. 42LT02, Sep. 2017, doi: 10.1088/1361-6528/aa897d.

[12] D. Buckley, N. C. G. Black, E. Castanon, C. Melios, M. Hardman, and O. Kazakova, "Frontiers of graphene and 2D material-based gas sensors for environmental monitoring," *2D Mater.*, 2020, doi: 10.1088/2053-1583/ab7bc5.

[13] D. Neumaier, S. Pindl, and M. C. Lemme, "Integrating graphene into semiconductor fabrication lines," *Nat. Mater.*, vol. 18, no. 6, pp. 525–529, Jun. 2019, doi: 10.1038/s41563-019-0359-7.

[14] D. Akinwande *et al.*, "Graphene and two-dimensional materials for silicon technology," *Nature*, vol. 573, no. 7775, pp. 507–518, Sep. 2019, doi: 10.1038/s41586-019-1573-9.

[15] C. Backes *et al.*, "Production and processing of graphene and related materials," *2D Mater.*, vol. 7, no. 2, p. 022001, Jan. 2020, doi: 10.1088/2053-1583/ab1e0a.

[16] H. Tomori *et al.*, "Introducing Nonuniform Strain to Graphene Using Dielectric Nanopillars," *Appl. Phys. Express*, vol. 4, no. 7, p. 075102, Jul. 2011, doi: 10.1143/APEX.4.075102.

[17] C. Lee, X. Wei, J. W. Kysar, and J. Hone, "Measurement of the Elastic Properties and Intrinsic Strength of Monolayer Graphene," *Science*, vol. 321, no. 5887, pp. 385–388, Jul. 2008, doi: 10.1126/science.1157996.

[18] A. S. Mayorov *et al.*, "Micrometer-Scale Ballistic Transport in Encapsulated Graphene at Room Temperature," *Nano Lett.*, vol. 11, no. 6, pp. 2396–2399, Jun. 2011, doi: 10.1021/nl200758b.

[19] R. R. Nair *et al.*, “Fine Structure Constant Defines Visual Transparency of Graphene,” *Science*, vol. 320, no. 5881, pp. 1308–1308, Jun. 2008, doi: 10.1126/science.1156965.
[20] J. S. Bunch *et al.*, “Impermeable Atomic Membranes from Graphene Sheets,” *Nano Lett.*, vol. 8, no. 8, pp. 2458–2462, Aug. 2008, doi: 10.1021/nl801457b.
[21] M. Lee *et al.*, “Sealing Graphene Nanodrums,” *Nano Lett.*, vol. 19, no. 8, pp. 5313–5318, Aug. 2019, doi: 10.1021/acs.nanolett.9b01770.
[22] P. Sun *et al.*, “How permeable is the impermeable graphene?,” *ArXiv191209220 Cond-Mat*, Dec. 2019, Accessed: Jan. 27, 2020. [Online]. Available: http://arxiv.org/abs/1912.09220.
[23] J. Moser, A. Barreiro, and A. Bachtold, “Current-induced cleaning of graphene,” *Appl. Phys. Lett.*, vol. 91, no. 16, p. 163513, Oct. 2007, doi: 10.1063/1.2789673.
[24] S. P. Koenig, N. G. Boddeti, M. L. Dunn, and J. S. Bunch, “Ultrastrong adhesion of graphene membranes,” *Nat. Nanotechnol.*, vol. 6, no. 9, pp. 543–546, Sep. 2011, doi: 10.1038/nnano.2011.123.
[25] A. D. Smith *et al.*, “Electromechanical Piezoresistive Sensing in Suspended Graphene Membranes,” *Nano Lett.*, vol. 13, no. 7, pp. 3237–3242, Jul. 2013, doi: 10.1021/nl401352k.
[26] M. R. Axet, O. Dechy-Cabaret, J. Durand, M. Gouygou, and P. Serp, “Coordination chemistry on carbon surfaces,” *Coord. Chem. Rev.*, vol. 308, pp. 236–345, Feb. 2016, doi: 10.1016/j.ccr.2015.06.005.
[27] K. S. Novoselov *et al.*, “Electric Field Effect in Atomically Thin Carbon Films,” *Science*, vol. 306, no. 5696, pp. 666–669, Oct. 2004, doi: 10.1126/science.1102896.
[28] C. R. Dean *et al.*, “Boron nitride substrates for high-quality graphene electronics,” *Nat. Nanotechnol.*, vol. 5, no. 10, pp. 722–726, Oct. 2010, doi: 10.1038/nnano.2010.172.
[29] C. Androulidakis, K. Zhang, M. Robertson, and S. Tawfick, “Tailoring the mechanical properties of 2D materials and heterostructures,” *2D Mater.*, vol. 5, no. 3, p. 032005, Jun. 2018, doi: 10.1088/2053-1583/aac764.
[30] S. Deng, L. Li, and Y. Zhang, “Strain Modulated Electronic, Mechanical, and Optical Properties of the Monolayer PdS2, PdSe2, and PtSe2 for Tunable Devices,” *ACS Appl. Nano Mater.*, vol. 1, no. 4, pp. 1932–1939, Apr. 2018, doi: 10.1021/acsanm.8b00363.
[31] S. Manzeli, A. Allain, A. Ghadimi, and A. Kis, “Piezoresistivity and Strain-induced Band Gap Tuning in Atomically Thin MoS2,” *Nano Lett.*, vol. 15, no. 8, pp. 5330–5335, Aug. 2015, doi: 10.1021/acs.nanolett.5b01689.
[32] M. Huang, T. A. Pascal, H. Kim, W. A. Goddard, and J. R. Greer, “Electronic−Mechanical Coupling in Graphene from in situ Nanoindentation Experiments and Multiscale Atomistic Simulations,” *Nano Lett.*, vol. 11, no. 3, pp. 1241–1246, Mar. 2011, doi: 10.1021/nl104227t.
[33] Y. Wang *et al.*, “Super-Elastic Graphene Ripples for Flexible Strain Sensors,” *ACS Nano*, vol. 5, no. 5, pp. 3645–3650, May 2011, doi: 10.1021/nn103523t.
[34] S.-E. Zhu, M. K. Ghatkesar, C. Zhang, and G. C. a. M. Janssen, “Graphene based piezoresistive pressure sensor,” *Appl. Phys. Lett.*, vol. 102, no. 16, p. 161904, Apr. 2013, doi: 10.1063/1.4802799.
[35] A. D. Smith *et al.*, “Piezoresistive Properties of Suspended Graphene Membranes under Uniaxial and Biaxial Strain in Nanoelectromechanical Pressure Sensors,” *ACS Nano*, vol. 10, no. 11, pp. 9879–9886, Nov. 2016, doi: 10.1021/acsnano.6b02533.
[36] R. Frisenda *et al.*, “Biaxial strain tuning of the optical properties of single-layer transition metal dichalcogenides,” *Npj 2D Mater. Appl.*, vol. 1, no. 1, p. 10, 2017.
[37] H.-Y. Qi, W.-T. Mi, H.-M. Zhao, T. Xue, Y. Yang, and T.-L. Ren, “A large-scale spray casting deposition method of WS2 films for high-sensitive, flexible and transparent sensor,” *Mater. Lett.*, vol. 201, pp. 161–164, Aug. 2017, doi: 10.1016/j.matlet.2017.04.062.

[38] G. López-Polín *et al.*, “Increasing the elastic modulus of graphene by controlled defect creation,” *Nat. Phys.*, vol. 11, no. 1, p. 26, 2015.
[39] H. Zhao, K. Min, and N. R. Aluru, “Size and Chirality Dependent Elastic Properties of Graphene Nanoribbons under Uniaxial Tension,” *Nano Lett.*, vol. 9, no. 8, pp. 3012–3015, Aug. 2009, doi: 10.1021/nl901448z.
[40] G. Kalosakas, N. N. Lathiotakis, C. Galiotis, and K. Papagelis, “In-plane force fields and elastic properties of graphene,” *J. Appl. Phys.*, vol. 113, no. 13, p. 134307, Apr. 2013, doi: 10.1063/1.4798384.
[41] O. L. Blakslee, D. G. Proctor, E. J. Seldin, G. B. Spence, and T. Weng, “Elastic Constants of Compression-Annealed Pyrolytic Graphite,” *J. Appl. Phys.*, vol. 41, no. 8, pp. 3373–3382, Jul. 1970, doi: 10.1063/1.1659428.
[42] M. Goldsche *et al.*, “Tailoring Mechanically Tunable Strain Fields in Graphene,” *Nano Lett.*, vol. 18, no. 3, pp. 1707–1713, Mar. 2018, doi: 10.1021/acs.nanolett.7b04774.
[43] K. I. Bolotin *et al.*, “Ultrahigh electron mobility in suspended graphene,” *Solid State Commun.*, vol. 146, no. 9–10, pp. 351–355, Jun. 2008, doi: 10.1016/j.ssc.2008.02.024.
[44] G.-H. Lee *et al.*, “High-Strength Chemical-Vapor–Deposited Graphene and Grain Boundaries,” *Science*, vol. 340, no. 6136, pp. 1073–1076, May 2013, doi: 10.1126/science.1235126.
[45] P. Zhang *et al.*, “Fracture toughness of graphene,” *Nat. Commun.*, vol. 5, no. 1, pp. 1–7, Apr. 2014, doi: 10.1038/ncomms4782.
[46] L. Banszerus *et al.*, “Ultrahigh-mobility graphene devices from chemical vapor deposition on reusable copper,” *Sci. Adv.*, vol. 1, no. 6, p. e1500222, Jul. 2015, doi: 10.1126/sciadv.1500222.
[47] L. Song *et al.*, “Large Scale Growth and Characterization of Atomic Hexagonal Boron Nitride Layers,” *Nano Lett.*, vol. 10, no. 8, pp. 3209–3215, Aug. 2010, doi: 10.1021/nl1022139.
[48] K. N. Kudin, G. E. Scuseria, and B. I. Yakobson, “C2F, BN, and C nanoshell elasticity from ab initio computations,” *Phys. Rev. B*, vol. 64, no. 23, p. 235406, Nov. 2001, doi: 10.1103/PhysRevB.64.235406.
[49] A. Falin *et al.*, “Mechanical properties of atomically thin boron nitride and the role of interlayer interactions,” *Nat. Commun.*, vol. 8, no. 1, pp. 1–9, Jun. 2017, doi: 10.1038/ncomms15815.
[50] Y. Kobayashi, C.-L. Tsai, and T. Akasaka, “Optical band gap of h-BN epitaxial film grown on c -plane sapphire substrate,” *Phys. Status Solidi C*, vol. 7, no. 7–8, pp. 1906–1908, 2010, doi: 10.1002/pssc.200983598.
[51] S. Bertolazzi, J. Brivio, and A. Kis, “Stretching and Breaking of Ultrathin MoS2,” *ACS Nano*, vol. 5, no. 12, pp. 9703–9709, Dec. 2011, doi: 10.1021/nn203879f.
[52] J. L. Feldman, “Elastic constants of 2H-MoS2 and 2H-NbSe2 extracted from measured dispersion curves and linear compressibilities,” *J. Phys. Chem. Solids*, vol. 37, no. 12, pp. 1141–1144, Jan. 1976, doi: 10.1016/0022-3697(76)90143-8.
[53] C. Poilane, P. Delobelle, C. Lexcellent, S. Hayashi, and H. Tobushi, “Analysis of the mechanical behavior of shape memory polymer membranes by nanoindentation, bulging and point membrane deflection tests,” *Thin Solid Films*, vol. 379, no. 1, pp. 156–165, Dec. 2000, doi: 10.1016/S0040-6090(00)01401-2.
[54] J. Wang *et al.*, “High Mobility MoS2 Transistor with Low Schottky Barrier Contact by Using Atomic Thick h-BN as a Tunneling Layer,” *Adv. Mater.*, vol. 28, no. 37, pp. 8302–8308, 2016, doi: 10.1002/adma.201602757.
[55] A. Splendiani *et al.*, “Emerging Photoluminescence in Monolayer MoS2,” *Nano Lett.*, vol. 10, no. 4, pp. 1271–1275, Apr. 2010, doi: 10.1021/nl903868w.

[56] K. F. Mak, C. Lee, J. Hone, J. Shan, and T. F. Heinz, "Atomically Thin MoS2: A New Direct-Gap Semiconductor," *Phys. Rev. Lett.*, vol. 105, no. 13, p. 136805, Sep. 2010, doi: 10.1103/PhysRevLett.105.136805.

[57] Y. Yang *et al.*, "Brittle Fracture of 2D MoSe2," *Adv. Mater.*, vol. 29, no. 2, p. 1604201, 2017, doi: 10.1002/adma.201604201.

[58] M. Hosseini, M. Elahi, M. Pourfath, and D. Esseni, "Very large strain gauges based on single layer MoSe2 and WSe2 for sensing applications," *Appl. Phys. Lett.*, vol. 107, no. 25, p. 253503, Dec. 2015, doi: 10.1063/1.4937438.

[59] P. Tonndorf *et al.*, "Photoluminescence emission and Raman response of monolayer $MoS_2$, $MoSe_2$, and $WSe_2$," *Opt. Express*, vol. 21, no. 4, pp. 4908–4916, Feb. 2013, doi: 10.1364/OE.21.004908.

[60] Y. Zhao *et al.*, "High-Electron-Mobility and Air-Stable 2D Layered PtSe2 FETs," *Adv. Mater.*, vol. 29, no. 5, p. 1604230, doi: 10.1002/adma.201604230.

[61] Y. Wang *et al.*, "Monolayer PtSe2, a New Semiconducting Transition-Metal-Dichalcogenide, Epitaxially Grown by Direct Selenization of Pt," *Nano Lett.*, vol. 15, no. 6, pp. 4013–4018, Jun. 2015, doi: 10.1021/acs.nanolett.5b00964.

[62] M. O'Brien *et al.*, "Raman characterization of platinum diselenide thin films," *2D Mater.*, vol. 3, no. 2, p. 021004, Apr. 2016, doi: 10.1088/2053-1583/3/2/021004.

[63] K. Liu *et al.*, "Elastic Properties of Chemical-Vapor-Deposited Monolayer MoS2, WS2, and Their Bilayer Heterostructures," *Nano Lett.*, vol. 14, no. 9, pp. 5097–5103, Sep. 2014, doi: 10.1021/nl501793a.

[64] F. Zeng, W.-B. Zhang, and B.-Y. Tang, "Electronic structures and elastic properties of monolayer and bilayer transition metal dichalcogenides MX2 (M = Mo, W; X = O, S, Se, Te): A comparative first-principles study," *Chin. Phys. B*, vol. 24, no. 9, p. 097103, Sep. 2015, doi: 10.1088/1674-1056/24/9/097103.

[65] M. W. Iqbal *et al.*, "High-mobility and air-stable single-layer $WS_2$ field-effect transistors sandwiched between chemical vapor deposition-grown hexagonal BN films," *Sci. Rep.*, vol. 5, p. 10699, Jun. 2015, doi: 10.1038/srep10699.

[66] W. Zhao *et al.*, "Origin of Indirect Optical Transitions in Few-Layer MoS2, WS2, and WSe2," *Nano Lett.*, vol. 13, no. 11, pp. 5627–5634, Nov. 2013, doi: 10.1021/nl403270k.

[67] R. Zhang, V. Koutsos, and R. Cheung, "Elastic properties of suspended multilayer WSe2," *Appl. Phys. Lett.*, vol. 108, no. 4, p. 042104, Jan. 2016, doi: 10.1063/1.4940982.

[68] J.-Y. Wang, Y. Li, Z.-Y. Zhan, T. Li, L. Zhen, and C.-Y. Xu, "Elastic properties of suspended black phosphorus nanosheets," *Appl. Phys. Lett.*, vol. 108, no. 1, p. 013104, Jan. 2016, doi: 10.1063/1.4939233.

[69] J. Qiao, X. Kong, Z.-X. Hu, F. Yang, and W. Ji, "High-mobility transport anisotropy and linear dichroism in few-layer black phosphorus," *Nat. Commun.*, vol. 5, no. 1, pp. 1–7, Jul. 2014, doi: 10.1038/ncomms5475.

[70] D. M. Warschauer, "Black Phosphorus as Strain Gauge and Pressure Transducer," *J. Appl. Phys.*, vol. 35, no. 12, pp. 3516–3519, Dec. 1964, doi: 10.1063/1.1713261.

[71] Z. Zhang *et al.*, "Strain-Modulated Bandgap and Piezo-Resistive Effect in Black Phosphorus Field-Effect Transistors," *Nano Lett.*, vol. 17, no. 10, pp. 6097–6103, Oct. 2017, doi: 10.1021/acs.nanolett.7b02624.

[72] X. Fan *et al.*, "Graphene ribbons with suspended masses as transducers in ultra-small nanoelectromechanical accelerometers," *Nat. Electron.*, pp. 1–11, Sep. 2019, doi: 10.1038/s41928-019-0287-1.

[73] C. S. Ruiz-Vargas *et al.*, "Softened Elastic Response and Unzipping in Chemical Vapor Deposition Graphene Membranes," *Nano Lett.*, vol. 11, no. 6, pp. 2259–2263, Jun. 2011, doi: 10.1021/nl200429f.

[74] M. Annamalai, S. Mathew, M. Jamali, D. Zhan, and M. Palaniapan, “Elastic and nonlinear response of nanomechanical graphene devices,” *J. Micromechanics Microengineering*, vol. 22, no. 10, p. 105024, Sep. 2012, doi: 10.1088/0960-1317/22/10/105024.
[75] M. Poot and H. S. J. van der Zant, “Nanomechanical properties of few-layer graphene membranes,” *Appl. Phys. Lett.*, vol. 92, no. 6, p. 063111, Feb. 2008, doi: 10.1063/1.2857472.
[76] D. Davidovikj, F. Alijani, S. J. Cartamil-Bueno, H. S. van der Zant, M. Amabili, and P. G. Steeneken, “Nonlinear dynamic characterization of two-dimensional materials,” *Nat. Commun.*, vol. 8, no. 1, p. 1253, 2017.
[77] R. N. Patel, J. P. Mathew, A. Borah, and M. M. Deshmukh, “Low tension graphene drums for electromechanical pressure sensing,” *2D Mater.*, vol. 3, no. 1, p. 011003, 2016, doi: 10.1088/2053-1583/3/1/011003.
[78] S. Lee *et al.*, “Electrically integrated SU-8 clamped graphene drum resonators for strain engineering,” *Appl. Phys. Lett.*, vol. 102, no. 15, p. 153101, Apr. 2013, doi: 10.1063/1.4793302.
[79] C. Gómez-Navarro, M. Burghard, and K. Kern, “Elastic Properties of Chemically Derived Single Graphene Sheets,” *Nano Lett.*, vol. 8, no. 7, pp. 2045–2049, Jul. 2008, doi: 10.1021/nl801384y.
[80] I. W. Frank, D. M. Tanenbaum, A. M. van der Zande, and P. L. McEuen, “Mechanical properties of suspended graphene sheets,” *J. Vac. Sci. Technol. B*, vol. 25, no. 6, pp. 2558–2561, Nov. 2007, doi: 10.1116/1.2789446.
[81] D. Garcia-Sanchez, A. M. van der Zande, A. S. Paulo, B. Lassagne, P. L. McEuen, and A. Bachtold, “Imaging Mechanical Vibrations in Suspended Graphene Sheets,” *Nano Lett.*, vol. 8, no. 5, pp. 1399–1403, May 2008, doi: 10.1021/nl080201h.
[82] A. M. van der Zande *et al.*, “Large-Scale Arrays of Single-Layer Graphene Resonators,” *Nano Lett.*, vol. 10, no. 12, pp. 4869–4873, Dec. 2010, doi: 10.1021/nl102713c.
[83] C. Chen *et al.*, “Performance of monolayer graphene nanomechanical resonators with electrical readout,” *Nat. Nanotechnol.*, vol. 4, no. 12, pp. 861–867, 2009, doi: 10.1038/nnano.2009.267.
[84] S. Wagner, C. Weisenstein, A. D. Smith, M. Östling, S. Kataria, and M. C. Lemme, “Graphene transfer methods for the fabrication of membrane-based NEMS devices,” *Microelectron. Eng.*, vol. 159, pp. 108–113, Jun. 2016, doi: 10.1016/j.mee.2016.02.065.
[85] C. Wirtz, N. C. Berner, and G. S. Duesberg, “Large-Scale Diffusion Barriers from CVD Grown Graphene,” *Adv. Mater. Interfaces*, vol. 2, no. 14, Sep. 2015, doi: 10.1002/admi.201500082.
[86] Q. Zhou and A. Zettl, “Electrostatic graphene loudspeaker,” *Appl. Phys. Lett.*, vol. 102, no. 22, p. 223109, Jun. 2013, doi: 10.1063/1.4806974.
[87] C. Berger, R. Phillips, A. Centeno, A. Zurutuza, and A. Vijayaraghavan, “Capacitive pressure sensing with suspended graphene–polymer heterostructure membranes,” *Nanoscale*, vol. 9, no. 44, pp. 17439–17449, 2017, doi: 10.1039/C7NR04621A.
[88] N. Iguiñiz, R. Frisenda, R. Bratschitsch, and A. Castellanos-Gomez, “Revisiting the Buckling Metrology Method to Determine the Young’s Modulus of 2D Materials,” *Adv. Mater.*, vol. 31, no. 10, p. 1807150, 2019, doi: 10.1002/adma.201807150.
[89] K. Parvez, S. Yang, X. Feng, and K. Müllen, “Exfoliation of graphene via wet chemical routes,” *Synth. Met.*, doi: 10.1016/j.synthmet.2015.07.014.
[90] J. A. Siddique, N. F. Attia, and K. E. Geckeler, “Polymer Nanoparticles as a tool for the exfoliation of graphene Sheets,” *Mater. Lett.*, doi: 10.1016/j.matlet.2015.05.134.
[91] J. H. Lee *et al.*, “One-Step Exfoliation Synthesis of Easily Soluble Graphite and Transparent Conducting Graphene Sheets,” *Adv. Mater.*, vol. 21, no. 43, pp. 4383–4387, 2009, doi: 10.1002/adma.200900726.

[92] K. R. Paton *et al.*, “Scalable production of large quantities of defect-free few-layer graphene by shear exfoliation in liquids,” *Nat. Mater.*, vol. 13, no. 6, pp. 624–630, Jun. 2014, doi: 10.1038/nmat3944.
[93] C. S. Boland *et al.*, “Sensitive electromechanical sensors using viscoelastic graphene-polymer nanocomposites,” *Science*, vol. 354, no. 6317, pp. 1257–1260, Dec. 2016, doi: 10.1126/science.aag2879.
[94] D. Hanlon *et al.*, “Liquid exfoliation of solvent-stabilized few-layer black phosphorus for applications beyond electronics,” *Nat. Commun.*, vol. 6, no. 1, pp. 1–11, Oct. 2015, doi: 10.1038/ncomms9563.
[95] Y. Gong *et al.*, “Layer-Controlled and Wafer-Scale Synthesis of Uniform and High-Quality Graphene Films on a Polycrystalline Nickel Catalyst,” *Adv. Funct. Mater.*, vol. 22, no. 15, pp. 3153–3159, 2012, doi: 10.1002/adfm.201200388.
[96] J. Jeon *et al.*, “Layer-controlled CVD growth of large-area two-dimensional MoS2 films,” *Nanoscale*, vol. 7, no. 5, pp. 1688–1695, Jan. 2015, doi: 10.1039/C4NR04532G.
[97] S. Kataria *et al.*, “Chemical vapor deposited graphene: From synthesis to applications,” *Phys. Status Solidi A*, vol. 211, no. 11, pp. 2439–2449, Nov. 2014, doi: 10.1002/pssa.201400049.
[98] T. Hallam, N. C. Berner, C. Yim, and G. S. Duesberg, “Strain, Bubbles, Dirt, and Folds: A Study of Graphene Polymer-Assisted Transfer,” *Adv. Mater. Interfaces*, vol. 1, no. 6, p. 1400115, 2014, doi: 10.1002/admi.201400115.
[99] A. Quellmalz *et al.*, “Wafer-Scale Transfer of Graphene by Adhesive Wafer Bonding,” in *2019 IEEE 32nd International Conference on Micro Electro Mechanical Systems (MEMS)*, Jan. 2019, pp. 257–259, doi: 10.1109/MEMSYS.2019.8870682.
[100] H.-S. Jang *et al.*, “Toward Scalable Growth for Single-Crystal Graphene on Polycrystalline Metal Foil,” *ACS Nano*, vol. 14, no. 3, pp. 3141–3149, Mar. 2020, doi: 10.1021/acsnano.9b08305.
[101] N. Mishra *et al.*, “Going beyond copper: wafer-scale synthesis of graphene on sapphire,” *Small*, vol. 15, no. 50, p. 1904906, Dec. 2019, doi: 10.1002/smll.201904906.
[102] Y. Wei, J. Wu, H. Yin, X. Shi, R. Yang, and M. Dresselhaus, “The nature of strength enhancement and weakening by pentagon–heptagon defects in graphene,” *Nat. Mater.*, vol. 11, no. 9, pp. 759–763, Sep. 2012, doi: 10.1038/nmat3370.
[103] K. S. Novoselov, A. Mishchenko, A. Carvalho, and A. H. C. Neto, “2D materials and van der Waals heterostructures,” *Science*, vol. 353, no. 6298, p. aac9439, Jul. 2016, doi: 10.1126/science.aac9439.
[104] Y. Kobayashi *et al.*, “Growth and Optical Properties of High-Quality Monolayer WS2 on Graphite,” *ACS Nano*, vol. 9, no. 4, pp. 4056–4063, Apr. 2015, doi: 10.1021/acsnano.5b00103.
[105] S. Kataria *et al.*, “Growth-Induced Strain in Chemical Vapor Deposited Monolayer MoS2: Experimental and Theoretical Investigation,” *Adv. Mater. Interfaces*, vol. 4, no. 17, p. 1700031, Sep. 2017, doi: 10.1002/admi.201700031.
[106] M. O’Brien *et al.*, “Transition Metal Dichalcogenide Growth via Close Proximity Precursor Supply,” *Sci. Rep.*, vol. 4, no. 1, pp. 1–7, Dec. 2014, doi: 10.1038/srep07374.
[107] S. Balendhran *et al.*, “Atomically thin layers of MoS2via a two step thermal evaporation–exfoliation method,” *Nanoscale*, vol. 4, no. 2, pp. 461–466, Jan. 2012, doi: 10.1039/C1NR10803D.
[108] Y. Zhan, Z. Liu, S. Najmaei, P. M. Ajayan, and J. Lou, “Large-area vapor-phase growth and characterization of MoS2 atomic layers on a SiO2 substrate,” *Small*, vol. 8, no. 7, pp. 966–971, Apr. 2012, doi: 10.1002/smll.201102654.

[109] S. Najmaei *et al.*, “Vapour phase growth and grain boundary structure of molybdenum disulphide atomic layers,” *Nat. Mater.*, vol. 12, no. 8, pp. 754–759, Aug. 2013, doi: 10.1038/nmat3673.

[110] R. Gatensby, T. Hallam, K. Lee, N. McEvoy, and G. S. Duesberg, “Investigations of vapour-phase deposited transition metal dichalcogenide films for future electronic applications,” *Solid-State Electron.*, vol. 125, pp. 39–51, Nov. 2016, doi: 10.1016/j.sse.2016.07.021.

[111] R. Gatensby *et al.*, “Controlled synthesis of transition metal dichalcogenide thin films for electronic applications,” *Appl. Surf. Sci.*, vol. 297, pp. 139–146, Apr. 2014, doi: 10.1016/j.apsusc.2014.01.103.

[112] K. Lee, R. Gatensby, N. McEvoy, T. Hallam, and G. S. Duesberg, “High-Performance Sensors Based on Molybdenum Disulfide Thin Films,” *Adv. Mater.*, vol. 25, no. 46, pp. 6699–6702, Dec. 2013, doi: 10.1002/adma.201303230.

[113] D. Kong *et al.*, “Synthesis of MoS2 and MoSe2 Films with Vertically Aligned Layers,” *Nano Lett.*, vol. 13, no. 3, pp. 1341–1347, Mar. 2013, doi: 10.1021/nl400258t.

[114] M. Naz *et al.*, “A New 2H-2H′/1T Cophase in Polycrystalline MoS2 and MoSe2 Thin Films,” *ACS Appl. Mater. Interfaces*, vol. 8, no. 45, pp. 31442–31448, Nov. 2016, doi: 10.1021/acsami.6b10972.

[115] M. Shanmugam, C. A. Durcan, and B. Yu, “Layered semiconductor molybdenum disulfide nanomembrane based Schottky-barrier solar cells,” *Nanoscale*, vol. 4, no. 23, p. 7399, 2012, doi: 10.1039/c2nr32394j.

[116] A. L. Elías *et al.*, “Controlled synthesis and transfer of large-area WS2 sheets: From single layer to few layers,” *ACS Nano*, vol. 7, no. 6, pp. 5235–5242, Jun. 2013, doi: 10.1021/nn400971k.

[117] M. O’Brien *et al.*, “Plasma assisted synthesis of WS2 for gas sensing applications,” *Chem. Phys. Lett.*, vol. 615, pp. 6–10, Nov. 2014, doi: 10.1016/j.cplett.2014.09.051.

[118] W. Kim *et al.*, “Field-Dependent Electrical and Thermal Transport in Polycrystalline WSe2,” *Adv. Mater. Interfaces*, vol. 5, no. 11, p. 1701161, 2018, doi: 10.1002/admi.201701161.

[119] C. Yim *et al.*, “High-Performance Hybrid Electronic Devices from Layered PtSe 2 Films Grown at Low Temperature,” *ACS Nano*, vol. 10, no. 10, pp. 9550–9558, 2016, doi: 10.1021/acsnano.6b04898.

[120] M. Yan *et al.*, “Lorentz-violating type-II Dirac fermions in transition metal dichalcogenide PtTe 2,” *Nat. Commun.*, vol. 8, no. 1, pp. 1–6, Aug. 2017, doi: 10.1038/s41467-017-00280-6.

[121] A. Reina *et al.*, “Large Area, Few-Layer Graphene Films on Arbitrary Substrates by Chemical Vapor Deposition,” *Nano Lett.*, vol. 9, no. 1, pp. 30–35, Jan. 2009, doi: 10.1021/nl801827v.

[122] J. W. Suk *et al.*, “Transfer of CVD-Grown Monolayer Graphene onto Arbitrary Substrates,” *ACS Nano*, vol. 5, no. 9, pp. 6916–6924, Sep. 2011, doi: 10.1021/nn201207c.

[123] C. J. L. de la Rosa *et al.*, “Frame assisted H2O electrolysis induced H2 bubbling transfer of large area graphene grown by chemical vapor deposition on Cu,” *Appl. Phys. Lett.*, vol. 102, no. 2, p. 022101, Jan. 2013, doi: 10.1063/1.4775583.

[124] A. Quellmalz *et al.*, “Large-Area Integration of 2D Material Heterostructures by Adhesive Bonding,” presented at the IEEE MEMS, Vancouver, Canada, 2020.

[125] A. Quellmalz *et al.*, “Large-Area Integration of Two-Dimensional Materials and Their Heterostructures by Wafer Bonding,” *Submitt. J. Publ.*, 2020.

[126] J. Kang, D. Shin, S. Bae, and B. H. Hong, “Graphene transfer: key for applications,” *Nanoscale*, vol. 4, no. 18, pp. 5527–5537, Aug. 2012, doi: 10.1039/C2NR31317K.

[127] S. J Cartamil-Bueno, A. Centeno, A. Zurutuza, P. Gerard Steeneken, H. S. J. van der Zant, and S. Houri, "Very large scale characterization of graphene mechanical devices using a colorimetry technique," *Nanoscale*, 2017, doi: 10.1039/C7NR01766A.

[128] R. J. Dolleman *et al.*, "Mass measurement of graphene using quartz crystal microbalances," *Appl. Phys. Lett.*, vol. 115, no. 5, p. 053102, Jul. 2019, doi: 10.1063/1.5111086.

[129] G. T. Kovacs, *Micromachined transducers sourcebook.* WCB/McGraw-Hill New York, 1998.

[130] T. Hallam, C. F. Moldovan, K. Gajewski, A. M. Ionescu, and G. S. Duesberg, "Large area suspended graphene for nano-mechanical devices," *Phys. Status Solidi B*, vol. 252, no. 11, pp. 2429–2432, Nov. 2015, doi: 10.1002/pssb.201552269.

[131] S. Vollebregt, R. J. Dolleman, H. S. J. van der Zant, P. G. Steeneken, and P. M. Sarro, "Suspended graphene beams with tunable gap for squeeze-film pressure sensing," in *2017 19th International Conference on Solid-State Sensors, Actuators and Microsystems (TRANSDUCERS)*, Jun. 2017, pp. 770–773, doi: 10.1109/TRANSDUCERS.2017.7994162.

[132] J. Romijn *et al.*, "A Miniaturized Low Power Pirani Pressure Sensor Based on Suspended Graphene," in *2018 IEEE 13th Annual International Conference on Nano/Micro Engineered and Molecular Systems (NEMS)*, Apr. 2018, pp. 11–14, doi: 10.1109/NEMS.2018.8556902.

[133] X. Fan *et al.*, "Suspended Graphene Membranes with Attached Silicon Proof Masses as Piezoresistive Nanoelectromechanical Systems Accelerometers," *Nano Lett.*, Sep. 2019, doi: 10.1021/acs.nanolett.9b01759.

[134] N. Tombros *et al.*, "Large yield production of high mobility freely suspended graphene electronic devices on a polydimethylglutarimide based organic polymer," *J. Appl. Phys.*, vol. 109, no. 9, p. 093702, May 2011, doi: 10.1063/1.3579997.

[135] Y. Oshidari, T. Hatakeyama, R. Kometani, S. Warisawa, and S. Ishihara, "High Quality Factor Graphene Resonator Fabrication Using Resist Shrinkage-Induced Strain," *Appl. Phys. Express*, vol. 5, no. 11, p. 117201, Oct. 2012, doi: 10.1143/APEX.5.117201.

[136] J. Sun, W. Wang, M. Muruganathan, and H. Mizuta, "Low pull-in voltage graphene electromechanical switch fabricated with a polymer sacrificial spacer," *Appl. Phys. Lett.*, vol. 105, no. 3, p. 033103, Jul. 2014, doi: 10.1063/1.4891055.

[137] O. I. Aydin, T. Hallam, J. L. Thomassin, M. Mouis, and G. S. Duesberg, "Interface and strain effects on the fabrication of suspended CVD graphene devices," *Solid-State Electron.*, vol. 108, pp. 75–83, Jun. 2015, doi: 10.1016/j.sse.2014.12.003.

[138] X. Fan *et al.*, "Manufacturing of Graphene Membranes with Suspended Silicon Proof Masses forMEMS and NEMS," *Microsyst. Nanoeng.*, no. in print, 2020, Accessed: Feb. 11, 2020. [Online]. Available: http://urn.kb.se/resolve?urn=urn:nbn:se:kth:diva-232551.

[139] A. D. Smith, S. Wagner, S. Kataria, B. G. Malm, M. C. Lemme, and M. Östling, "Wafer-Scale Statistical Analysis of Graphene FETs Part I: Wafer-Scale Fabrication and Yield Analysis," *IEEE Trans. Electron Devices*, vol. 64, no. 9, pp. 3919–3926, Sep. 2017, doi: 10.1109/TED.2017.2727820.

[140] A. D. Smith, S. Wagner, S. Kataria, B. G. Malm, M. C. Lemme, and M. Östling, "Wafer-Scale Statistical Analysis of Graphene Field-Effect Transistors Part II: Analysis of Device Properties," *IEEE Trans. Electron Devices*, vol. 64, no. 9, pp. 3927–3933, Sep. 2017, doi: 10.1109/TED.2017.2727823.

[141] P. Li, G. Jing, B. Zhang, S. Sando, and T. Cui, "Wafer-size free-standing single-crystalline graphene device arrays," *Appl. Phys. Lett.*, vol. 105, no. 8, p. 083118, Aug. 2014, doi: 10.1063/1.4894255.

[142] Y. Lee *et al.*, "Wafer-Scale Synthesis and Transfer of Graphene Films," *Nano Lett.*, vol. 10, no. 2, pp. 490–493, Feb. 2010, doi: 10.1021/nl903272n.

[143] S. Rahimi *et al.*, "Toward 300 mm Wafer-Scalable High-Performance Polycrystalline Chemical Vapor Deposited Graphene Transistors," *ACS Nano*, vol. 8, no. 10, pp. 10471–10479, Oct. 2014, doi: 10.1021/nn5038493.

[144] C. S. Boland *et al.*, "PtSe2 grown directly on polymer foil for use as a robust piezoresistive sensor," *2D Mater.*, vol. 6, no. 4, p. 045029, Aug. 2019, doi: 10.1088/2053-1583/ab33a1.

[145] S. Wagner *et al.*, "Noninvasive Scanning Raman Spectroscopy and Tomography for Graphene Membrane Characterization," *Nano Lett.*, vol. 17, no. 3, pp. 1504–1511, Mar. 2017, doi: 10.1021/acs.nanolett.6b04546.

[146] S. Wittmann, C. Glacer, S. Wagner, S. Pindl, and M. C. Lemme, "Graphene Membranes for Hall Sensors and Microphones Integrated with CMOS-Compatible Processes," *ACS Appl. Nano Mater.*, vol. 2, no. 8, pp. 5079–5085, Aug. 2019, doi: 10.1021/acsanm.9b00998.

[147] A. Castellanos-Gomez, V. Singh, H. S. J. van der Zant, and G. A. Steele, "Mechanics of freely-suspended ultrathin layered materials," *Ann. Phys.*, vol. 527, no. 1–2, pp. 27–44, Jan. 2015, doi: 10.1002/andp.201400153.

[148] D. Davidovikj, J. J. Slim, S. J. Cartamil-Bueno, H. S. J. van der Zant, P. G. Steeneken, and W. J. Venstra, "Visualizing the Motion of Graphene Nanodrums," *Nano Lett.*, vol. 16, no. 4, pp. 2768–2773, Apr. 2016, doi: 10.1021/acs.nanolett.6b00477.

[149] S. J. Cartamil-Bueno, P. G. Steeneken, A. Centeno, A. Zurutuza, H. S. J. van der Zant, and S. Houri, "Colorimetry Technique for Scalable Characterization of Suspended Graphene," *Nano Lett.*, vol. 16, no. 11, pp. 6792–6796, Nov. 2016, doi: 10.1021/acs.nanolett.6b02416.

[150] A. D. Smith *et al.*, "Resistive graphene humidity sensors with rapid and direct electrical readout," *Nanoscale*, Nov. 2015, doi: 10.1039/C5NR06038A.

[151] A. D. Smith *et al.*, "Graphene-based CO 2 sensing and its cross-sensitivity with humidity," *RSC Adv.*, vol. 7, no. 36, pp. 22329–22339, 2017, doi: 10.1039/C7RA02821K.

[152] A. Ghosh, D. J. Late, L. S. Panchakarla, A. Govindaraj, and C. N. R. Rao, "NO2 and humidity sensing characteristics of few-layer graphenes," *J. Exp. Nanosci.*, vol. 4, no. 4, pp. 313–322, Dec. 2009, doi: 10.1080/17458080903115379.

[153] A. Quellmalz *et al.*, "Influence of Humidity on Contact Resistance in Graphene Devices," *ACS Appl. Mater. Interfaces*, vol. 10, no. 48, pp. 41738–41746, Dec. 2018, doi: 10.1021/acsami.8b10033.

[154] S. Riazimehr *et al.*, "Spectral sensitivity of graphene/silicon heterojunction photodetectors," *Solid-State Electron.*, vol. 115, Part B, pp. 207–212, Jan. 2016, doi: 10.1016/j.sse.2015.08.023.

[155] L. Britnell *et al.*, "Strong Light-Matter Interactions in Heterostructures of Atomically Thin Films," *Science*, vol. 340, no. 6138, pp. 1311–1314, Jun. 2013, doi: 10.1126/science.1235547.

[156] X. Fan *et al.*, "Humidity and CO2 gas sensing properties of double-layer graphene," *Carbon*, Nov. 2017, doi: 10.1016/j.carbon.2017.11.038.

[157] J. T. Smith, A. D. Franklin, D. B. Farmer, and C. D. Dimitrakopoulos, "Reducing Contact Resistance in Graphene Devices through Contact Area Patterning," *ACS Nano*, vol. 7, no. 4, pp. 3661–3667, Apr. 2013, doi: 10.1021/nn400671z.

[158] F. Giubileo and A. Di Bartolomeo, "The role of contact resistance in graphene field-effect devices," *Prog. Surf. Sci.*, vol. 92, no. 3, pp. 143–175, Aug. 2017, doi: 10.1016/j.progsurf.2017.05.002.

[159] T. Cusati *et al.*, "Electrical properties of graphene-metal contacts," *Sci. Rep.*, vol. 7, Jul. 2017, doi: 10.1038/s41598-017-05069-7.

[160] L. Anzi *et al.*, "Ultra-low contact resistance in graphene devices at the Dirac point," *2D Mater.*, vol. 5, no. 2, p. 025014, 2018, doi: 10.1088/2053-1583/aaab96.

[161] V. Passi *et al.*, “Ultralow Specific Contact Resistivity in Metal–Graphene Junctions via Contact Engineering,” *Adv. Mater. Interfaces*, vol. 6, no. 1, p. 1801285, 2019, doi: 10.1002/admi.201801285.

[162] Z. Cheng *et al.*, “Immunity to Contact Scaling in MoS2 Transistors Using in Situ Edge Contacts,” *Nano Lett.*, vol. 19, no. 8, pp. 5077–5085, Aug. 2019, doi: 10.1021/acs.nanolett.9b01355.

[163] T. G. Beckwith, R. D. Marangoni, and J. H. V. Lienhard, “Mechanical measurements,” *CERN Document Server*, 2009. https://cds.cern.ch/record/1394311 (accessed Feb. 10, 2020).

[164] A. L. Window, *Strain gauge technology*. London; New York: Elsevier Applied Science, 1992.

[165] S. Yang and N. Lu, “Gauge Factor and Stretchability of Silicon-on-Polymer Strain Gauges,” *Sensors*, vol. 13, no. 7, pp. 8577–8594, Jul. 2013, doi: 10.3390/s130708577.

[166] S.-H. Bae, Y. Lee, B. K. Sharma, H.-J. Lee, J.-H. Kim, and J.-H. Ahn, “Graphene-based transparent strain sensor,” *Carbon*, vol. 51, pp. 236–242, Jan. 2013, doi: 10.1016/j.carbon.2012.08.048.

[167] A. Tarasov, M.-Y. Tsai, H. Taghinejad, P. M. Campbell, A. Adibi, and E. M. Vogel, “Piezoresistive strain sensing with flexible MoS2 field-effect transistors,” in *Device Research Conference (DRC), 2015 73rd Annual*, Jun. 2015, pp. 159–160, doi: 10.1109/DRC.2015.7175604.

[168] J. T. M. van Beek *et al.*, “Scalable 1.1 GHz fundamental mode piezo-resistive silicon MEMS resonator,” in *2007 IEEE International Electron Devices Meeting*, Dec. 2007, pp. 411–414, doi: 10.1109/IEDM.2007.4418960.

[169] M. Sansa, M. Fernández-Regúlez, J. Llobet, Á. San Paulo, and F. Pérez-Murano, “High-sensitivity linear piezoresistive transduction for nanomechanical beam resonators,” *Nat. Commun.*, vol. 5, p. 4313, 2014.

[170] J. Xia, F. Chen, J. Li, and N. Tao, “Measurement of the quantum capacitance of graphene,” *Nat. Nanotechnol.*, vol. 4, no. 8, p. 505, 2009.

[171] C. Chen *et al.*, “Graphene mechanical oscillators with tunable frequency,” *Nat. Nanotechnol.*, vol. 8, no. 12, p. 923, 2013.

[172] J. Zhang, Y. Zhao, Y. Ge, M. Li, L. Yang, and X. Mao, “Design Optimization and Fabrication of High-Sensitivity SOI Pressure Sensors with High Signal-to-Noise Ratios Based on Silicon Nanowire Piezoresistors,” *Micromachines*, vol. 7, no. 10, p. 187, 2016, doi: 10.3390/mi7100187.

[173] L. Kumar *et al.*, “MEMS oscillating squeeze-film pressure sensor with optoelectronic feedback,” *J. Micromechanics Microengineering*, vol. 25, no. 4, p. 045011, Mar. 2015, doi: 10.1088/0960-1317/25/4/045011.

[174] K. E. Wojciechowski, B. E. Boser, and A. P. Pisano, “A MEMS resonant strain sensor with 33 nano-strain resolution in a 10 kHz bandwidth,” in *2005 IEEE SENSORS*, Oct. 2005, pp. 4 pp.-, doi: 10.1109/ICSENS.2005.1597857.

[175] R. J. Dolleman, D. Davidovikj, S. J. Cartamil-Bueno, H. S. J. van der Zant, and P. G. Steeneken, “Graphene Squeeze-Film Pressure Sensors,” *Nano Lett.*, vol. 16, no. 1, pp. 568–571, Jan. 2016, doi: 10.1021/acs.nanolett.5b04251.

[176] D. Davidovikj, M. Poot, S. J. Cartamil-Bueno, H. S. J. van der Zant, and P. G. Steeneken, “On-chip Heaters for Tension Tuning of Graphene Nanodrums,” *Nano Lett.*, vol. 18, no. 5, pp. 2852–2858, May 2018, doi: 10.1021/acs.nanolett.7b05358.

[177] T. B. Gabrielson, “Mechanical-thermal noise in micromachined acoustic and vibration sensors,” *IEEE Trans. Electron Devices*, vol. 40, no. 5, pp. 903–909, May 1993, doi: 10.1109/16.210197.

[178] R. J. Dolleman, S. Houri, A. Chandrashekar, F. Alijani, H. S. J. van der Zant, and P. G. Steeneken, "Opto-thermally excited multimode parametric resonance in graphene membranes," *Sci. Rep.*, vol. 8, no. 1, pp. 1–7, Jun. 2018, doi: 10.1038/s41598-018-27561-4.
[179] D. Davidovikj, D. Bouwmeester, H. S. J. van der Zant, and P. G. Steeneken, "Graphene gas pumps," *2D Mater.*, vol. 5, no. 3, p. 031009, May 2018, doi: 10.1088/2053-1583/aac0a8.
[180] R. Singh, R. J. T. Nicholl, K. I. Bolotin, and S. Ghosh, "Motion Transduction with Thermo-mechanically Squeezed Graphene Resonator Modes," *Nano Lett.*, vol. 18, no. 11, pp. 6719–6724, Nov. 2018, doi: 10.1021/acs.nanolett.8b02293.
[181] G. J. Verbiest, J. N. Kirchhof, J. Sonntag, M. Goldsche, T. Khodkov, and C. Stampfer, "Detecting Ultrasound Vibrations with Graphene Resonators," *Nano Lett.*, vol. 18, no. 8, pp. 5132–5137, Aug. 2018, doi: 10.1021/acs.nanolett.8b02036.
[182] D. Davidovikj, P. H. Scheepers, H. S. J. van der Zant, and P. G. Steeneken, "Static Capacitive Pressure Sensing Using a Single Graphene Drum," *ACS Appl. Mater. Interfaces*, vol. 9, no. 49, pp. 43205–43210, Dec. 2017, doi: 10.1021/acsami.7b17487.
[184] O. N. Tufte, P. W. Chapman, and D. Long, "Silicon Diffused-Element Piezoresistive Diaphragms," *J. Appl. Phys.*, vol. 33, no. 11, pp. 3322–3327, Nov. 1962, doi: 10.1063/1.1931164.
[185] Murata Electronics Oy, "Capacitive absolute 1.2 bar SCB10H-B012FB pressure sensor element," *SCB10H-B012FB datasheet*. https://www.murata.com/~/media/webrenewal/products/sensor/gyro/element/pressure/datasheet_scb10h_.ashx?la=en (accessed Nov. 23, 2018).
[186] Q. Wang, W. Hong, and L. Dong, "Graphene 'microdrums' on a freestanding perforated thin membrane for high sensitivity MEMS pressure sensors," *Nanoscale*, 2016, doi: 10.1039/C5NR09274D.
[187] J. Aguilera-Servin, T. Miao, and M. Bockrath, "Nanoscale pressure sensors realized from suspended graphene membrane devices," *Appl. Phys. Lett.*, vol. 106, no. 8, p. 083103, Feb. 2015, doi: 10.1063/1.4908176.
[188] A. D. Smith *et al.*, "Electromechanical Piezoresistive Sensing in Suspended Graphene Membranes," *Nano Lett.*, vol. 13, no. 7, pp. 3237–3242, Jul. 2013, doi: 10.1021/nl401352k.
[189] A. Dehe, K. Fricke, K. Mutamba, and H. L. Hartnagel, "A piezoresistive GaAs pressure sensor with GaAs/AlGaAs membrane technology," *J. Micromechanics Microengineering*, vol. 5, no. 2, p. 139, 1995, doi: 10.1088/0960-1317/5/2/021.
[190] C. Stampfer *et al.*, "Fabrication of Single-Walled Carbon-Nanotube-Based Pressure Sensors," *Nano Lett.*, vol. 6, no. 2, pp. 233–237, Feb. 2006, doi: 10.1021/nl052171d.
[191] C. K. M. Fung, M. Q. H. Zhang, R. H. M. Chan, and W. J. Li, "A PMMA-based micro pressure sensor chip using carbon nanotubes as sensing elements," in *18th IEEE International Conference on Micro Electro Mechanical Systems, 2005. MEMS 2005*, Jan. 2005, pp. 251–254, doi: 10.1109/MEMSYS.2005.1453914.
[192] J. H. Kim, K. T. Park, H. C. Kim, and K. Chun, "Fabrication of a piezoresistive pressure sensor for enhancing sensitivity using silicon nanowire," in *TRANSDUCERS 2009 - 2009 International Solid-State Sensors, Actuators and Microsystems Conference*, Jun. 2009, pp. 1936–1939, doi: 10.1109/SENSOR.2009.5285668.
[193] R. He and P. Yang, "Giant piezoresistance effect in silicon nanowires," *Nat. Nanotechnol.*, vol. 1, no. 1, p. nnano.2006.53, Oct. 2006, doi: 10.1038/nnano.2006.53.
[194] G. J. Verbiest *et al.*, "Integrated impedance bridge for absolute capacitance measurements at cryogenic temperatures and finite magnetic fields," *Rev. Sci. Instrum.*, vol. 90, no. 8, p. 084706, Aug. 2019, doi: 10.1063/1.5089207.
[195] M. Šiškins *et al.*, "Sensitive capacitive pressure sensors based on graphene membrane arrays," *ArXiv200308869 Cond-Mat Physicsphysics*, Mar. 2020, Accessed: Mar. 24, 2020. [Online]. Available: http://arxiv.org/abs/2003.08869.

[196] Y.-M. Chen *et al.*, “Ultra-large suspended graphene as highly elastic membrane for capacitive pressure sensor,” *Nanoscale*, Jan. 2016, doi: 10.1039/C5NR08668J.
[197] J. Lee, Z. Wang, K. He, J. Shan, and P. X.-L. Feng, “Air damping of atomically thin MoS2 nanomechanical resonators,” *Appl. Phys. Lett.*, vol. 105, no. 2, p. 023104, Jul. 2014, doi: 10.1063/1.4890387.
[198] R. J. T. Nicholl, N. V. Lavrik, I. Vlassiouk, B. R. Srijanto, and K. I. Bolotin, “Hidden Area and Mechanical Nonlinearities in Freestanding Graphene,” *Phys. Rev. Lett.*, vol. 118, no. 26, p. 266101, Jun. 2017, doi: 10.1103/PhysRevLett.118.266101.
[199] R. Puers, S. Reyntjens, and D. De Bruyker, “The NanoPirani—an extremely miniaturized pressure sensor fabricated by focused ion beam rapid prototyping,” *Sens. Actuators Phys.*, vol. 97–98, pp. 208–214, Apr. 2002, doi: 10.1016/S0924-4247(01)00863-9.
[200] R. R. Spender, B. M. Fleischer, P. W. Barth, and J. B. Angell, “A theoretical study of transducer noise in piezoresistive and capacitive silicon pressure sensors,” *IEEE Trans. Electron Devices*, vol. 35, no. 8, pp. 1289–1298, Aug. 1988, doi: 10.1109/16.2550.
[201] K. L. Ekinci, Y. T. Yang, and M. L. Roukes, “Ultimate limits to inertial mass sensing based upon nanoelectromechanical systems,” *J. Appl. Phys.*, vol. 95, no. 5, pp. 2682–2689, Feb. 2004, doi: 10.1063/1.1642738.
[202] TDK Electronics, “EPCOS Product Profile 2018 Pressure Sensor Dies,” *Sensors - Pressure Sensors Dies*. https://en.tdk-electronics.tdk.com/download/174158/b9cfd64b38cbeaebdd11ff32a17fe894/pressure-sensors-dies-pp.pdf (accessed Nov. 23, 2018).
[203] N. A. Hall, M. Okandan, R. Littrell, B. Bicen, and F. L. Degertekin, “Micromachined optical microphone structures with low thermal-mechanical noise levels,” *J. Acoust. Soc. Am.*, vol. 122, no. 4, pp. 2031–2037, Sep. 2007, doi: 10.1121/1.2769615.
[204] Q. Zhou, J. Zheng, S. Onishi, M. F. Crommie, and A. K. Zettl, “Graphene electrostatic microphone and ultrasonic radio,” *Proc. Natl. Acad. Sci.*, vol. 112, no. 29, pp. 8942–8946, Jul. 2015, doi: 10.1073/pnas.1505800112.
[205] D. Todorović *et al.*, “Multilayer graphene condenser microphone,” *2D Mater.*, vol. 2, no. 4, p. 045013, Nov. 2015, doi: 10.1088/2053-1583/2/4/045013.
[206] Infineon Technologies, “High performance digital XENSIVTM MEMS microphone IM69D130.” IM69D130 datasheet, Accessed: Jan. 28, 2020. [Online]. Available: https://www.infineon.com/dgdl/Infineon-IM69D130-DS-v01_00-EN.pdf?fileId=5546d462602a9dc801607a0e46511a2e.
[207] M. Fueldner and A. Dehé, “Dual back plate silicon MEMS microphone: Balancing high performance,” *DAGA 2015 41 Jahrestag. Für Akust. Nürnberg Ger.*, 2015.
[208] G. S. Wood *et al.*, “Design and Characterization of a Micro-Fabricated Graphene-Based MEMS Microphone,” *IEEE Sens. J.*, vol. 19, no. 17, pp. 7234–7242, Sep. 2019, doi: 10.1109/JSEN.2019.2914401.
[209] A. Laitinen *et al.*, “A graphene resonator as an ultrasound detector for generalized Love waves in a polymer film with two level states,” *J. Phys. Appl. Phys.*, vol. 52, no. 24, p. 24LT02, Apr. 2019, doi: 10.1088/1361-6463/ab11a9.
[210] K. Matsui *et al.*, “Mechanical properties of few layer graphene cantilever,” in *2014 IEEE 27th International Conference on Micro Electro Mechanical Systems (MEMS)*, Jan. 2014, pp. 1087–1090, doi: 10.1109/MEMSYS.2014.6765834.
[211] M. K. Blees *et al.*, “Graphene kirigami,” *Nature*, vol. 524, no. 7564, p. 204, 2015.
[212] A. M. Hurst, S. Lee, W. Cha, and J. Hone, “A graphene accelerometer,” in *2015 28th IEEE International Conference on Micro Electro Mechanical Systems (MEMS)*, Jan. 2015, pp. 865–868, doi: 10.1109/MEMSYS.2015.7051096.

[213] J. W. Kang, J. H. Lee, H. J. Hwang, and K.-S. Kim, "Developing accelerometer based on graphene nanoribbon resonators," *Phys. Lett. A*, vol. 376, no. 45, pp. 3248–3255, Oct. 2012, doi: 10.1016/j.physleta.2012.08.040.

[214] F.-T. Shi, S.-C. Fan, C. Li, and X.-B. Peng, "Modeling and Analysis of a Novel Ultrasensitive Differential Resonant Graphene Micro-Accelerometer with Wide Measurement Range," *Sensors*, vol. 18, no. 7, p. 2266, Jul. 2018, doi: 10.3390/s18072266.

[215] K.-R. Byun, K.-S. Kim, H. J. Hwang, and J. W. Kang, "Sensitivity of Graphene-Nanoribbon-Based Accelerometer with Attached Mass," *J. Comput. Theor. Nanosci.*, vol. 10, no. 8, pp. 1886–1891, Aug. 2013, doi: 10.1166/jctn.2013.3144.

[216] X. Du, I. Skachko, A. Barker, and E. Y. Andrei, "Approaching ballistic transport in suspended graphene," *Nat. Nanotechnol.*, vol. 3, no. 8, pp. 491–495, Aug. 2008, doi: 10.1038/nnano.2008.199.

[217] K. I. Bolotin, K. J. Sikes, J. Hone, H. L. Stormer, and P. Kim, "Temperature-Dependent Transport in Suspended Graphene," *Phys. Rev. Lett.*, vol. 101, no. 9, p. 096802, Aug. 2008, doi: 10.1103/PhysRevLett.101.096802.

[218] E. V. Castro *et al.*, "Limits on Charge Carrier Mobility in Suspended Graphene due to Flexural Phonons," *Phys. Rev. Lett.*, vol. 105, no. 26, p. 266601, Dec. 2010, doi: 10.1103/PhysRevLett.105.266601.

[219] S. V. Morozov *et al.*, "Giant Intrinsic Carrier Mobilities in Graphene and Its Bilayer," *Phys. Rev. Lett.*, vol. 100, no. 1, p. 016602, Jan. 2008, doi: 10.1103/PhysRevLett.100.016602.

[220] Z. Wang, M. Shaygan, M. Otto, D. Schall, and D. Neumaier, "Flexible Hall sensors based on graphene," *Nanoscale*, vol. 8, no. 14, pp. 7683–7687, 2016, doi: 10.1039/C5NR08729E.

[221] N. D. Arora, J. R. Hauser, and D. J. Roulston, "Electron and hole mobilities in silicon as a function of concentration and temperature," *IEEE Trans. Electron Devices*, vol. 29, no. 2, pp. 292–295, Feb. 1982, doi: 10.1109/T-ED.1982.20698.

[222] H. Xu *et al.*, "Batch-fabricated high-performance graphene Hall elements," *Sci. Rep.*, vol. 3, p. 1207, Feb. 2013, doi: 10.1038/srep01207.

[223] A. Dankert, B. Karpiak, and S. P. Dash, "Hall sensors batch-fabricated on all-CVD h-BN/graphene/h-BN heterostructures," *Sci. Rep.*, vol. 7, no. 1, pp. 1–7, Nov. 2017, doi: 10.1038/s41598-017-12277-8.

[224] Infineon Technologies, "Linear Hall IC TLE4997A8." TLE4997A8 datasheet, Accessed: Jan. 26, 2020. [Online]. Available: https://www.infineon.com/dgdl/Infineon-TLE4997A8D_DS-DS-v01_01-EN.pdf?fileId=5546d4625b62cd8a015bc87f823f319f.

[225] L. Banszerus *et al.*, "Extraordinary high room-temperature carrier mobility in graphene-WSe2 heterostructures," *ArXiv190909523 Cond-Mat*, Sep. 2019, Accessed: Oct. 07, 2019. [Online]. Available: http://arxiv.org/abs/1909.09523.

[226] L. Wang *et al.*, "One-Dimensional Electrical Contact to a Two-Dimensional Material," *Science*, vol. 342, no. 6158, pp. 614–617, Nov. 2013, doi: 10.1126/science.1244358.

[227] J. Dauber *et al.*, "Ultra-sensitive Hall sensors based on graphene encapsulated in hexagonal boron nitride," *Appl. Phys. Lett.*, vol. 106, no. 19, p. 193501, May 2015, doi: 10.1063/1.4919897.

[228] M. Gautam and A. H. Jayatissa, "Graphene based field effect transistor for the detection of ammonia," *J. Appl. Phys.*, vol. 112, no. 6, p. 064304, Sep. 2012, doi: 10.1063/1.4752272.

[229] E. J. Olson *et al.*, "Capacitive Sensing of Intercalated H2O Molecules Using Graphene," *ACS Appl. Mater. Interfaces*, vol. 7, no. 46, pp. 25804–25812, Nov. 2015, doi: 10.1021/acsami.5b07731.

[230] M.-C. Chen, C.-L. Hsu, and T.-J. Hsueh, "Fabrication of Humidity Sensor Based on Bilayer Graphene," *IEEE Electron Device Lett.*, vol. 35, no. 5, pp. 590–592, May 2014, doi: 10.1109/LED.2014.2310741.

[231] N. Lei, P. Li, W. Xue, and J. Xu, "Simple graphene chemiresistors as pH sensors: fabrication and characterization," *Meas. Sci. Technol.*, vol. 22, no. 10, p. 107002, 2011, doi: 10.1088/0957-0233/22/10/107002.

[232] M. Myers, J. Cooper, B. Pejcic, M. Baker, B. Raguse, and L. Wieczorek, "Functionalized graphene as an aqueous phase chemiresistor sensing material," *Sens. Actuators B Chem.*, vol. 155, no. 1, pp. 154–158, 2011, doi: 10.1016/j.snb.2010.11.040.

[233] N. Ruecha, R. Rangkupan, N. Rodthongkum, and O. Chailapakul, "Novel paper-based cholesterol biosensor using graphene/polyvinylpyrrolidone/polyaniline nanocomposite," *Biosens. Bioelectron.*, vol. 52, pp. 13–19, 2014, doi: 10.1016/j.bios.2013.08.018.

[234] F.-L. Meng, Z. Guo, and X.-J. Huang, "Graphene-based hybrids for chemiresistive gas sensors," *TrAC Trends Anal. Chem.*, vol. 68, pp. 37–47, 2015, doi: 10.1016/j.trac.2015.02.008.

[235] S. Kumar, S. Kaushik, R. Pratap, and S. Raghavan, "Graphene on Paper: A Simple, Low-Cost Chemical Sensing Platform," *ACS Appl. Mater. Interfaces*, vol. 7, no. 4, pp. 2189–2194, 2015, doi: 10.1021/am5084122.

[236] Y. H. Y.-J. J. Kim *et al.*, "Self-Activated Transparent All-Graphene Gas Sensor with Endurance to Humidity and Mechanical Bending," *ACS Nano*, vol. 9, no. 10, pp. 10453–10460, 2015, doi: 10.1021/acsnano.5b04680.

[237] M. Singh *et al.*, "Noncovalently Functionalized Monolayer Graphene for Sensitivity Enhancement of Surface Plasmon Resonance Immunosensors," *J. Am. Chem. Soc.*, vol. 137, no. 8, pp. 2800–2803, Mar. 2015, doi: 10.1021/ja511512m.

[238] Y. Dan, Y. Lu, N. J. Kybert, Z. Luo, and A. T. C. Johnson, "Intrinsic Response of Graphene Vapor Sensors," *Nano Lett.*, vol. 9, no. 4, pp. 1472–1475, Apr. 2009, doi: 10.1021/nl8033637.

[239] Y. Chen *et al.*, "Electronic Detection of Lectins Using Carbohydrate-Functionalized Nanostructures: Graphene versus Carbon Nanotubes," *ACS Nano*, vol. 6, no. 1, pp. 760–770, Jan. 2012, doi: 10.1021/nn2042384.

[240] S. Rumyantsev, G. Liu, W. Stillman, M. Shur, and A. A. Balandin, "Electrical and noise characteristics of graphene field-effect transistors: ambient effects, noise sources and physical mechanisms," *J. Phys. Condens. Matter*, vol. 22, no. 39, p. 395302, Oct. 2010, doi: 10.1088/0953-8984/22/39/395302.

[241] A. A. Balandin, "Low-frequency 1/f noise in graphene devices," *Nat. Nanotechnol.*, vol. 8, no. 8, pp. 549–555, Aug. 2013, doi: 10.1038/nnano.2013.144.

[242] Renesas, "IDT SGAS707." IDT SGAS707 datasheet, Accessed: Jan. 27, 2020. [Online]. Available: https://www.idt.com/eu/en/document/dst/sgas707.

[243] S.-J. Choi and I.-D. Kim, "Recent Developments in 2D Nanomaterials for Chemiresistive-Type Gas Sensors," *Electron. Mater. Lett.*, vol. 14, no. 3, pp. 221–260, May 2018, doi: 10.1007/s13391-018-0044-z.

[244] G. Yang, C. Lee, J. Kim, F. Ren, and S. J. Pearton, "Flexible graphene-based chemical sensors on paper substrates," *Phys. Chem. Chem. Phys.*, vol. 15, no. 6, pp. 1798–1801, Jan. 2013, doi: 10.1039/C2CP43717A.

[245] S.-Y. Cho *et al.*, "Highly Enhanced Gas Adsorption Properties in Vertically Aligned MoS2 Layers," *ACS Nano*, vol. 9, no. 9, pp. 9314–9321, Sep. 2015, doi: 10.1021/acsnano.5b04504.

[246] J. Sun, M. Muruganathan, and H. Mizuta, "Room temperature detection of individual molecular physisorption using suspended bilayer graphene," *Sci. Adv.*, vol. 2, no. 4, Apr. 2016, doi: 10.1126/sciadv.1501518.

[247] AMS, "AMS AS-MLV-P2 Air Quality Sensor." AS-MLV-P2 datasheet, Accessed: Jan. 28, 2020. [Online]. Available: https://ams.com/documents/20143/36005/AS-MLV-P2_DS000359_1-00.pdf/43d38978-7af6-ed5c-f1dd-46950855abca.

[248] A. De Luca *et al.*, “Temperature-modulated graphene oxide resistive humidity sensor for indoor air quality monitoring,” *Nanoscale*, vol. 8, no. 8, pp. 4565–4572, 2016, doi: 10.1039/C5NR08598E.
[249] H.-Y. Kim, K. Lee, N. McEvoy, C. Yim, and G. S. Duesberg, “Chemically Modulated Graphene Diodes,” *Nano Lett.*, vol. 13, no. 5, pp. 2182–2188, May 2013, doi: 10.1021/nl400674k.
[250] A. Singh, M. A. Uddin, T. Sudarshan, and G. Koley, “Tunable Reverse-Biased Graphene/Silicon Heterojunction Schottky Diode Sensor,” *Small*, vol. 10, no. 8, pp. 1555–1565, Apr. 2014, doi: 10.1002/smll.201302818.
[251] H. Li *et al.*, “Fabrication of Single- and Multilayer MoS2 Film-Based Field-Effect Transistors for Sensing NO at Room Temperature,” *Small*, vol. 8, no. 1, pp. 63–67, 2012, doi: 10.1002/smll.201101016.
[252] D. J. Late, T. Doneux, and M. Bougouma, “Single-layer MoSe2 based NH3 gas sensor,” *Appl. Phys. Lett.*, vol. 105, no. 23, p. 233103, Dec. 2014, doi: 10.1063/1.4903358.
[253] I. Shackery *et al.*, “Few-layered α-MoTe 2 Schottky junction for a high sensitivity chemical-vapour sensor,” *J. Mater. Chem. C*, vol. 6, no. 40, pp. 10714–10722, 2018, doi: 10.1039/C8TC02635A.
[254] Y. Kim *et al.*, “Two-Dimensional NbS2 Gas Sensors for Selective and Reversible NO2 Detection at Room Temperature,” *ACS Sens.*, vol. 4, no. 9, pp. 2395–2402, Sep. 2019, doi: 10.1021/acssensors.9b00992.
[255] A. Yang *et al.*, “Humidity sensing using vertically oriented arrays of ReS2nanosheets deposited on an interdigitated gold electrode,” *2D Mater.*, vol. 3, no. 4, p. 045012, Oct. 2016, doi: 10.1088/2053-1583/3/4/045012.
[256] M. Sajjad, E. Montes, N. Singh, and U. Schwingenschlögl, “Superior Gas Sensing Properties of Monolayer PtSe2,” *Adv. Mater. Interfaces*, vol. 4, no. 5, p. 1600911, 2017, doi: 10.1002/admi.201600911.
[257] M. Kodu *et al.*, “Graphene-Based Ammonia Sensors Functionalised with Sub-Monolayer V2O5: A Comparative Study of Chemical Vapour Deposited and Epitaxial Graphene †,” *Sensors*, vol. 19, no. 4, p. 951, Jan. 2019, doi: 10.3390/s19040951.
[258] S. P. Koenig, L. Wang, J. Pellegrino, and J. S. Bunch, “Selective molecular sieving through porous graphene,” *Nat. Nanotechnol.*, vol. 7, no. 11, p. 728, 2012.
[259] H. Li *et al.*, “Ultrathin, molecular-sieving graphene oxide membranes for selective hydrogen separation,” *Science*, vol. 342, no. 6154, pp. 95–98, 2013.
[260] R. Joshi *et al.*, “Precise and ultrafast molecular sieving through graphene oxide membranes,” *science*, vol. 343, no. 6172, pp. 752–754, 2014.
[261] R. J. Dolleman, S. J. Cartamil-Bueno, H. S. J. van der Zant, and P. G. Steeneken, “Graphene gas osmometers,” *2D Mater.*, vol. 4, no. 1, p. 011002, 2017, doi: 10.1088/2053-1583/4/1/011002.
[262] I. E. Rosłoń *et al.*, “Graphene Effusion-based Gas Sensor,” *ArXiv200109509 Cond-Mat Physicsphysics*, Jan. 2020, Accessed: Jan. 28, 2020. [Online]. Available: http://arxiv.org/abs/2001.09509.
[263] J. Chaste, A. Eichler, J. Moser, G. Ceballos, R. Rurali, and A. Bachtold, “A nanomechanical mass sensor with yoctogram resolution,” *Nat. Nanotechnol.*, vol. 7, no. 5, pp. 301–304, May 2012, doi: 10.1038/nnano.2012.42.
[264] Y. T. Yang, C. Callegari, X. L. Feng, K. L. Ekinci, and M. L. Roukes, “Zeptogram-Scale Nanomechanical Mass Sensing,” *Nano Lett.*, vol. 6, no. 4, pp. 583–586, Apr. 2006, doi: 10.1021/nl052134m.
[265] H.-L. Lee, Y.-C. Yang, and W.-J. Chang, “Mass Detection Using a Graphene-Based Nanomechanical Resonator,” *Jpn. J. Appl. Phys.*, vol. 52, no. 2R, p. 025101, Jan. 2013, doi: 10.7567/JJAP.52.025101.

[266] I.-B. Baek *et al.*, “Attogram mass sensing based on silicon microbeam resonators,” *Sci. Rep.*, vol. 7, p. 46660, Apr. 2017, doi: 10.1038/srep46660.
[267] INFICON Inc., “Research Quartz Crystal Microbalance IPN 603800 Rev.K.” IPN 603800 Rev.K Operation and Service Manual, Accessed: Jan. 27, 2020. [Online]. Available: https://products.inficon.com/getattachment.axd/?attaName=6538a92e-efcf-4dc2-ba00-1297c15b938e.
[268] A. Blaikie, D. Miller, and B. J. Alemán, “A fast and sensitive room-temperature graphene nanomechanical bolometer,” *Nat. Commun.*, vol. 10, no. 1, pp. 1–8, Oct. 2019, doi: 10.1038/s41467-019-12562-2.
[269] T. Endoh *et al.*, “Uncooled infrared detector with 12um pixel pitch video graphics array,” in *Infrared Technology and Applications XXXIX*, Jun. 2013, vol. 8704, p. 87041G, doi: 10.1117/12.2013690.
[270] U. Mizrahi *et al.*, “Large-format 17μm high-end VOx μ-bolometer infrared detector,” in *Infrared Technology and Applications XXXIX*, Jun. 2013, vol. 8704, p. 87041H, doi: 10.1117/12.2015653.
[271] H.-H. Yang and G. M. Rebeiz, “Sub-10-pW/Hz0.5 Uncooled Micro-Bolometer With a Vacuum Micro-Package,” *IEEE Trans. Microw. Theory Tech.*, vol. 64, no. 7, pp. 2129–2136, Jul. 2016, doi: 10.1109/TMTT.2016.2562623.
[272] G. D. Skidmore, C. J. Han, and C. Li, “Uncooled microbolometers at DRS and elsewhere through 2013,” in *Image Sensing Technologies: Materials, Devices, Systems, and Applications*, May 2014, vol. 9100, p. 910003, doi: 10.1117/12.2054135.
[273] A. Rogalski, P. Martyniuk, and M. Kopytko, “Challenges of small-pixel infrared detectors: a review,” *Rep. Prog. Phys.*, vol. 79, no. 4, p. 046501, Mar. 2016, doi: 10.1088/0034-4885/79/4/046501.
[274] D. Davidovikj *et al.*, “Ultrathin complex oxide nanomechanical resonators,” *ArXiv190500056 Cond-Mat Physicsphysics*, Apr. 2019, Accessed: Jan. 22, 2020. [Online]. Available: http://arxiv.org/abs/1905.00056.
[275] J. López-Cabrelles *et al.*, “Isoreticular two-dimensional magnetic coordination polymers prepared through pre-synthetic ligand functionalization,” *Nat. Chem.*, vol. 10, no. 10, pp. 1001–1007, Oct. 2018, doi: 10.1038/s41557-018-0113-9.
[276] S. Milana, “The lab-to-fab journey of 2D materials,” *Nat. Nanotechnol.*, vol. 14, no. 10, pp. 919–921, Oct. 2019, doi: 10.1038/s41565-019-0554-3.